\documentclass{aa}  

\usepackage{graphicx}
\usepackage[varg]{txfonts}
\usepackage{amsmath}
\usepackage{amssymb}
\usepackage[none]{hyphenat}
\usepackage{enumitem}
\usepackage{hyperref}

\hypersetup{
    colorlinks=true,    
    linkcolor=blue,
    citecolor=blue,
    urlcolor=blue,
}

\begin{document} 

   \title{Bounds on galaxy stochasticity from halo occupation\\distribution modeling}

   \subtitle{}

   \author{Dylan Britt\inst{1,2,3,4}\fnmsep\thanks{djbritt@stanford.edu (DB)}
          \and
          Daniel Gruen\inst{4,5}
          \and
          Oliver Friedrich\inst{4,5}
          \and
          Sihan Yuan\inst{2,3}
          \and
          Bernardita Ried Guachalla\inst{1,2,3,4}
    }

    \institute{
            Department of Physics, Stanford University, 382 Via Pueblo Mall, Stanford, CA 94305, USA
         \and
             Kavli Institute for Particle Astrophysics \& Cosmology, 452 Lomita Mall, Stanford, CA 94305, USA
        \and
            SLAC National Accelerator Laboratory, 2575 Sand Hill Road, Menlo Park, CA 94025, USA
        \and
            University Observatory, Faculty of Physics, Ludwig-Maximilians-Universit\"{a}t, Scheinerstra{\ss}e 1, 81679 Munich, Germany
        \and
            Excellence Cluster ORIGINS, Boltzmannstra{\ss}e 2, 85748 Garching, Germany
    }

   \date{Received XXX; accepted YYY}

    \abstract
    {The joint probability distribution of matter overdensity and galaxy counts in cells is a powerful probe of cosmology, and the extent to which variance in galaxy counts at fixed matter density deviates from Poisson shot noise is not fully understood. The lack of informed bounds on this stochasticity is currently the limiting factor in constraining cosmology with the galaxy-matter PDF. We investigate stochasticity in the conditional distribution of galaxy counts along lines of sight with fixed matter density, and we present a halo occupation distribution (HOD)-based approach for obtaining plausible ranges for stochasticity parameters. To probe the high-dimensional space of possible galaxy-matter connections, we derive a set of HODs which conserve the galaxies' linear bias and number density to produce \textsc{redMaGiC}-like galaxy catalogs within the \textsc{AbacusSummit} suite of $N$-body simulations. We study the impact of individual HOD parameters and cosmology on stochasticity and perform a Monte Carlo search in HOD parameter space, subject to the constraints on bias and density. In mock catalogs generated by the selected HODs, shot noise in galaxy counts spans both sub-Poisson and super-Poisson values, ranging from 80\% to 133\% of Poisson variance for cells with mean matter density. Nearly all of the derived HODs show a positive relationship between local matter density and stochasticity. For galaxy catalogs with higher stochasticity, modeling galaxy bias to second order is required for an accurate description of the conditional PDF of galaxy counts at fixed matter density. The presence of galaxy assembly bias also substantially extends the range of stochasticity in the super-Poisson direction. This HOD-based approach leverages degrees of freedom in the galaxy-halo connection to obtain informed bounds on model nuisance parameters, and it can be adapted to study other parametrizations of shot noise in galaxy counts, in particular to motivate prior ranges on stochasticity for cosmological analyses.}


    \maketitle


\section{Introduction}

Understanding the statistical connection between galaxies and dark matter is essential to inferring the physics of the Universe from galaxy surveys. In the standard $\Lambda$CDM model of cosmology, the present large-scale structure is dominated by cold dark matter concentrated in gravitationally bound halos \citep{White_Rees,Cooray}, with a constant vacuum energy density driving the acceleration of cosmic expansion \citep{weinberg2013}. The ability to probe this dark sector hinges on models that relate observable tracers, among them galaxies, to properties of the dark matter structures they inhabit (see \citealt{wechsler_tinker} for a comprehensive review). Several complementary cosmological probes require reliable models of the galaxy-matter connection, in particular galaxy clustering \citep{BOSS_DR12_clustering,DESY1_clusters}, weak gravitational lensing \citep{Tyson,Hoekstra,Mandelbaum,KIDS1000_shear,DESY3_shear}, and their cross-correlations \citep[e.g.][]{KIDS1000_3x2pt,DESY3_3x2pt_main,DESY3_3x2pt_redmagic,DESY3_3x2pt_maglim}.

As the volume and precision of galaxy surveys increase, so must the accuracy of models of the joint distribution of galaxies and matter, if systematic modeling errors are to remain below the level of statistical uncertainties and observational systematics. Current and upcoming surveys such as the Dark Energy Spectroscopic Instrument (DESI; \citealt{desicollaboration2016desi}), the Euclid space telescope \citep{Euclid}, and the Vera C. Rubin Observatory's Legacy Survey of Space and Time (LSST; \citealt{Ivezic_2019}) will surpass their predecessors by roughly an order of magnitude in terms of total galaxies cataloged. The associated increase in statistical power promises a major improvement in cosmological constraining power, provided that the models of the tracer-matter connection employed in cosmological analyses do not dominate their uncertainty budgets. Achieving this goal requires striking a balance between systematic errors that dominate when a model is too simple and statistical errors incurred when a model is overly flexible.

Developing reliable models directly from theory proves challenging, as both the gravitational collapse of dark matter into halos and the assembly of baryonic matter into galaxies are substantially nonlinear. On physical scales for which the cosmic density field is linear or quasi-linear, perturbative galaxy bias expansions provide a rigorous and well-studied description of the statistical connection between galaxy density and matter density (see \citealt{Desjacques} for a review). These expansions inevitably break down on smaller scales, including nonlinear regimes probed by observables such as galaxy clustering and galaxy-galaxy lensing.

A number of empirically-motivated approaches address the need to model the galaxy-matter connection on nonlinear scales, including halo occupation distributions (HODs, \citealt{Peacock_Smith,Berlind}), which analytically relate expected galaxy counts to properties of host dark matter halos and introduce additional free parameters which must be constrained by observations. Galaxies and matter may also be modeled in terms of the joint distribution of their counts (in the case of galaxies) or density fields, in 2D or 3D, without the need to choose a specific halo definition. This approach appears in various counts-in-cells statistics for galaxies and matter either separately or jointly, and these have proven useful for tasks such as constraining cosmology \citep[e.g.][]{Bel,Codis2016,Uhlemann2017velocity,Uhlemann2017Hunting,Uhlemann2018,Friedrich_2018,Gruen_2018,Friedrich2020,Repp,Uhlemann2020,Gough,Burger2023} and measuring galaxy bias parameters \citep{Uhlemann2017,Wang}. In the broader context of probes of large-scale structure, these one-point PDFs are complementary to standard two-point correlation functions \citep[e.g.][]{Bautista_eBOSS,Krause2021,Joachimi_KiDS,Li_HSC2023} and higher-order statistics of the galaxy and matter density fields such as three-point functions \citep{Schneider_3pt,Takada_3pt}, integrated three-point functions \citep{Chiang_2014,Halder2021,Halder_2023,Halder2022,Gong2023}, and nearest-neighbor distributions \citep{Banerjee_Abel2020,Banerjee_Abel2021,Wang_kNN2022,Yuan_kNN2023,Yuan_2DkNN}.

One obstacle to using one-point statistics in cosmological analyses is that accurate modeling of the galaxy-matter PDF may require the introduction of nuisance parameters which are neither well-constrained by existing observations nor well-described theoretically. This was the case for the Dark Energy Survey (DES) Year 1 density split statistics analysis \citep{Friedrich_2018,Gruen_2018}, which found the need for two additional parameters describing non-Poisson shot noise in galaxy counts. The lack of well-motivated bounds on these stochasticity parameters necessitated wide prior distributions which dominated the uncertainty budget of the density split analysis. A related agnostic approach to nonlinear galaxy bias and stochasticity is a limiting factor also for cosmological analyses of lensing and galaxy clustering two-point correlation functions \citep{Pandey_2020,Sugiyama2020}.

For statistics of the large-scale galaxy distribution such as galaxy stochasticity, HODs provide a useful forward model for establishing well-motivated bounds or prior distributions. In its most general form, an HOD makes only the weak assumption that all galaxies are situated in dark matter halos and that the properties and formation history of a halo determine the number and properties of the galaxies it contains. If one can identify the subspace of HOD parameters which matches some chosen summary statistics of the galaxy catalog of interest, then these HODs can be used to generate mock galaxy catalogs in $N$-body simulations, from which measurements of stochasticity or any other nuisance parameter may be made directly.

In this work we obtain such a set of HODs with the aim of probing the parameter space corresponding to a luminous red galaxy (LRG) sample selected by the \textsc{redMaGiC} algorithm \citep{Rozo_redmagic}. By producing mock catalogs with these HODs and measuring their stochasticity, we aim to set a plausible range on the parametrization of \citet{Friedrich_2018} and \citet{Gruen_2018}. A companion paper (Ried Guachalla et al., in prep.) presents a methodology for using these ranges to obtain prior distributions for stochasticity parameters which minimize error in a frequentist interpretation of the marginalized posterior distributions over cosmological parameters of interest.

The structure of this paper is as follows: in Sect.~\ref{sec:bias}, we introduce the framework used to study galaxy bias and stochasticity in our analyses and review HODs and their extension to include galaxy assembly bias. In Sect.~\ref{sec:methods}, we describe the rationale and computational methods (including simulations) used to select a set of realistic HODs, as well as the methods used to compute galaxy counts and matter densities in cells and measure bias and stochasticity. Section~\ref{sec:results} presents our results, including the selected HODs, their stochasticity, and the effects of changing cosmology and cell geometry. In Sect.~\ref{sec:disc}, we interpret our findings regarding the relationship between the galaxy-matter connection and stochasticity, and we discuss the implications of our results for modeling the galaxy-matter PDF and placing priors on stochasticity parameters. We conclude in Sect.~\ref{sec:concl} with a brief outlook on applying informed bounds on galaxy stochasticity in cosmological analyses.


\section{Modeling the galaxy-matter connection}
\label{sec:bias}

\subsection{Stochasticity in galaxy counts}
\label{subsec:stochasticity}
We consider the analysis of survey data consisting of galaxy counts $N_\mathrm{g}$ and matter overdensities $\delta_\mathrm{m}$ along different lines of sight. Throughout this work we assume a circular top-hat filter applied along each line of sight; such a filter is defined by an angular scale and redshift range, in the case of a galaxy survey, or a comoving radius and cylinder depth, in the case of a simulation snapshot. Each line of sight $i$ thus corresponds to a comoving volume (cell) containing $N_{\mathrm{g},i}$ galaxies and a mean matter overdensity $\delta_{\mathrm{m},i}$. In galaxy survey applications, matter density cannot be measured directly and must be replaced by an observational proxy such as gravitational shear. Selecting cells of a given matter overdensity yields the discrete probability distribution $P(N_\mathrm{g\,}\vert\,\delta_\mathrm{m})$, which contains information on the galaxy-matter connection and is part of several related probes of cosmology \citep{Gruen_trough,Friedrich_2018,Gruen_2018,Uhlemann2020,Friedrich_pdf}. A simple approach to modeling this distribution is to assume that it is approximately Poisson, with an expected galaxy count that varies linearly with the chosen matter density:
\begin{equation}
    \label{eq:linbias}
    P(N_\mathrm{g\,}\vert\,\delta_\mathrm{m})\,=\,\exp\left[-\bar{N}_\mathrm{g}(1+b\delta_\mathrm{m})\right]\times\left[\bar{N}_\mathrm{g}(1+b\delta_\mathrm{m})\right]^{\,N_\mathrm{g}}\times\left(N_\mathrm{g}!\right)^{-1}.
\end{equation}
Here $\bar{N}_\mathrm{g}$ is the mean galaxy count across all cells (of all matter densities), and $b$ is the linear galaxy bias. 

On physical scales for which galaxy counts in 3D are approximately Poisson-distributed and galaxies are linearly biased with respect to matter, galaxies will remain Poisson-distributed and linearly biased in a 2D projection as well. However, the assumption of strictly linear bias inevitably breaks down on sufficiently small scales, and there is also no a priori reason that the distribution must be exactly Poisson. Indeed, the density split analysis of \citet{Gruen_2018} demonstrated that the mean and variance of $P(N_\mathrm{g\,}\vert\,\delta_\mathrm{m})$ are not equal, as they would be for a simple Poisson distribution. Relative to the model of equation \ref{eq:linbias}, \citet{Friedrich_2018} and \citet{Gruen_2018} found improved fits to simulations and survey data using a generalization of the Poisson distribution given by
\begin{align}
    \label{eq:stoch}
    P(N_{\mathrm{g}\,}\vert\,\delta_\mathrm{m})\,=\,\frac{1}{\mathcal{N}}\,\exp\left(-\,\frac{\bar{N}_\mathrm{g}(1+b\delta_\mathrm{m})}{\alpha(\delta_\mathrm{m})}\right)&\times\left(\frac{\bar{N}_\mathrm{g}(1+b\delta_\mathrm{m})}{\alpha(\delta_\mathrm{m})}\right)^{N_\mathrm{g}/\alpha(\delta_\mathrm{m})}\notag\\[0.5em]
    &\hspace{0.5em}\times\left[\,\Gamma\left(\frac{N_\mathrm{g}}{\alpha(\delta_\mathrm{m})}+1\right)\,\right]^{\,-1},
\end{align}
where $\Gamma$ is the Gamma function and \mbox{$\alpha(\delta_\mathrm{m})\equiv\alpha_0+\alpha_1\delta_\mathrm{m}$} modulates the variance of the distribution in the following manner:
\begin{align*}
    \alpha<1\ &:\ \text{sub-Poisson}\\
    \alpha=1\ &:\ \text{Poisson}\\
    \alpha>1\ &:\ \text{super-Poisson\,.}
\end{align*}
This distribution can be interpreted as one in which galaxies appear in sets of $\alpha$, with the sets themselves exhibiting simple Poisson variance. Note that the probability on the right side of equation~\ref{eq:stoch} corresponds to drawing a (generally non-integer) value $N_\mathrm{g}/\alpha$, not the value $N_\mathrm{g}$ itself. This requires a continuous analogue of the Poisson distribution, which is achieved by replacing the factorial in the Poisson probability with the Gamma function; this still yields a valid probability mass function over integer values of $N_\mathrm{g}$. The distribution is normalized by \mbox{$\mathcal{N}\simeq1/\alpha$}, where the equality is exact for integer values of $\alpha$ \citep{Friedrich_2018} and close to exact for all $\alpha$ given a sufficiently large expectation value for $P(N_\mathrm{g\,}\vert\,\delta_\mathrm{m})$, as is the case in this work.

In this parametrization, non-Poisson shot noise in galaxy counts (hereafter \textit{stochasticity}) consists of a global component $\alpha_0$ that affects all cells equally and a parameter $\alpha_1$ that captures the dependence of stochasticity on local matter density. Pathological cases do arise when $\alpha(\delta_\mathrm{m})$ and/or the quantity \mbox{$1+b\delta_\mathrm{m}$} are nonpositive; the numerical handling of such cells in our analysis is described in Sect.~\ref{subsubsec:cics_stoch}.

The HODs used in this work are derived to match specific values of galaxy density and linear bias as defined in equation \ref{eq:stoch}, the same model applied to galaxy counts in cells and cosmic shear in \citet{Friedrich_2018} and \citet{Gruen_2018}. However, in the mock catalogs produced by these HODs, linear galaxy bias alone is insufficient to fully describe $P(N_\mathrm{g\,}\vert\,\delta_\mathrm{m})$ for some of the galaxy-matter connections probed at larger matter under- or overdensities. When analyzing stochasticity in the mock catalogs, we therefore model galaxy bias to second order by modifying equation \ref{eq:stoch} according to
\begin{align}
   \label{eq:quadbias}
   1+b\delta_\mathrm{m}\ \rightarrow\ 1+b_1\delta_\mathrm{m}+\frac{b_2}{2}\left(\delta_\mathrm{m}^{\,2}-\sigma_\mathrm{m}^{\,2}\right)\,,
\end{align}
where $b_1$ and $b_2$ now denote the linear and quadratic galaxy bias, respectively, and \mbox{$\sigma_\mathrm{m}^{\,2}\equiv\langle\delta_\mathrm{m}^{\,2}\rangle$} is the variance of matter overdensity across all cells. The full model for the conditional distribution of galaxy counts at fixed matter density, including quadratic bias, is then
\begin{align}
    \label{eq:stoch_quad}
    P(N_\mathrm{g\,}\vert\,\delta_\mathrm{m})=\,&\frac{1}{\alpha_0+\alpha_1\delta_\mathrm{m}}\,\exp\left(-\,\frac{\bar{N}_\mathrm{g}\,\Big[1+b_1\delta_\mathrm{m}+\frac{b_2}{2}\big(\delta_\mathrm{m}^{\,2}-\sigma_\mathrm{m}^{\,2}\big)\Big]}{\alpha_0+\alpha_1\delta_\mathrm{m}}\right)\notag\\[0.5em]
    &\hspace{1.5em}\times\,\left(\frac{\bar{N}_\mathrm{g}\,\Big[1+b_1\delta_\mathrm{m}+\frac{b_2}{2}\big(\delta_\mathrm{m}^{\,2}-\sigma_\mathrm{m}^{\,2}\big)\Big]}{\alpha_0+\alpha_1\delta_\mathrm{m}}\right)^{N_\mathrm{g}/(\alpha_0+\alpha_1\delta_\mathrm{m})}\notag\\[0.5em]
    &\hspace{4em}\times\,\left[\,\Gamma\left(\frac{N_\mathrm{g}}{\alpha_0+\alpha_1\delta_\mathrm{m}}+1\right)\,\right]^{\,-1},
\end{align}
where the $1/\alpha(\delta_\mathrm{m})$ normalization has been assumed. As a standard caution, we note that linear bias denoted by $b$ or $b_1$ is not an equivalent quantity across different models such as equations \ref{eq:linbias}, \ref{eq:stoch}, and \ref{eq:stoch_quad}. In the remainder of this work, ``bias'' and ``linear bias'' refer to $b_1$ as defined in the stochasticity model of equation~\ref{eq:stoch_quad}, unless otherwise specified.

Galaxy stochasticity in the $\alpha_0,\alpha_1$ parametrization is not well understood from a physical perspective, nor has it been explored in depth via simulations. The density split statistics analysis of \citet{Gruen_2018} used conservatively wide prior distributions for stochasticity parameters, with the authors noting that the cosmological constraining power of the model would improve significantly with even modestly tightened priors. In this work, we aim to better understand the plausible range of $\alpha_0$ and $\alpha_1$ and to develop a method for deriving this range given basic properties of a galaxy sample, such as its linear bias and galaxy number density. A companion paper (Ried Guachalla et al., in prep.) presents a selection algorithm for obtaining informative priors on nuisance parameters such as $\alpha_0$ and $\alpha_1$, based on such ranges. These priors are optimized in the sense that the resulting marginalized 1D posteriors for cosmological parameters are minimally biased, i.e. such that prior volume effects are minimized. These informed, total-error-minimizing (ITEM) priors enable one to assess the improvement in cosmological constraining power associated with a tightening of the plausible ranges for $\alpha_0$ and $\alpha_1$ via the methods described in Sect.~\ref{sec:methods}.

\subsection{Example: modeling the counts-in-cells PDF}
\label{subsec:example_pdf}

\begin{figure*}
    \centering
    \resizebox{\hsize}{!}
    {\includegraphics{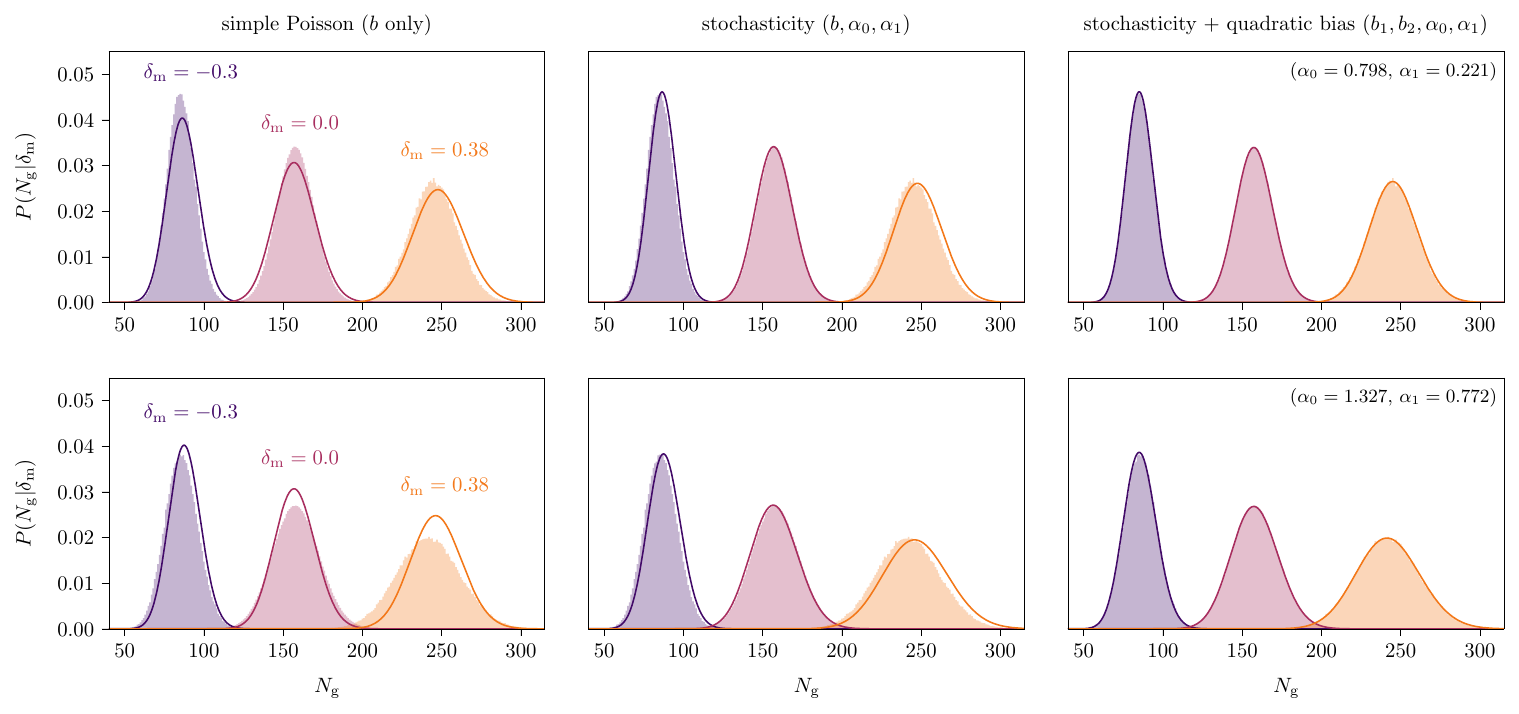}}
    \caption{Comparison of different model predictions for $P(N_\mathrm{g\,}\vert\,\delta_\mathrm{m})$ against measured counts-in-cells distributions in mock galaxy catalogs for a pair of example HODs which have sub-Poisson scatter (top row) and super-Poisson scatter (bottom row) at \mbox{$\delta_\mathrm{m}=0$}. Within each row, the three panels show model fits (solid curves) for the different parametrizations of $P(N_\mathrm{g\,}\vert\,\delta_\mathrm{m})$ given in equations~\ref{eq:linbias}, \ref{eq:stoch}, and \ref{eq:stoch_quad}. Left column: best fitting model with linear bias only, i.e. a simple Poisson distribution. Middle column: model with linear bias $b$ and density-dependent stochasticity $\alpha_0,\alpha_1$, which decouples the mean and variance of the distribution. Right column: stochasticity model with galaxy bias modeled to second order in $\delta_\mathrm{m}$. Shaded histograms show the measured distributions of galaxy counts in cells for three narrow bins of width 0.05 in $\delta_\mathrm{m}$. Each histogram is an average over 200 mock galaxy catalogs. The underdense ($-0.3$) and overdense ($0.38$) bins are set to the most extreme values of $\delta_\mathrm{m}$ for which each bin still contains at least 1000 of the 160,000 cylinders in the simulation box. Stochasticity values $\alpha_0$ and $\alpha_1$ for the two HODs (as measured in the model with quadratic bias) are given in the top right corner of the last panel in each row.}
    \label{fig:pdfs}
\end{figure*}

To illustrate the role of stochasticity and bias parameters in modeling $P(N_\mathrm{g\,}\vert\,\delta_\mathrm{m})$, in Fig.~\ref{fig:pdfs} we fit three different models to measured counts-in-cells distributions for mock galaxies. The models are fit to two different sets of mock catalogs generated by different HODs, one with sub-Poisson scatter and one with super-Poisson scatter at \mbox{$\delta_\mathrm{m}=0$} (i.e. \mbox{$\alpha_0<1$} and \mbox{$\alpha_0>1$}, respectively).\footnote{These HODs are the two with the most extreme values of $\alpha_0$ in the results of our Monte Carlo search, described in Sect.~\ref{subsec:mc_hods} and shown in Fig.~\ref{fig:stoch_mc}.} Fig.~\ref{fig:pdfs} shows the measured counts-in-cells distributions at three different matter overdensities \mbox{$\delta_\mathrm{m}\in\{-0.3,\,0.0,\,0.38\}$}, along with model fits for the PDFs defined in equations~\ref{eq:linbias}, \ref{eq:stoch}, and \ref{eq:stoch_quad}. The two nonzero values of $\delta_\mathrm{m}$ are set at the most extreme values for which there exist at least 1000 cylinders whose matter overdensities fall in a narrow bin of width $0.05$ in $\delta_\mathrm{m}$. For a given bin, each model is evaluated at each individual cylinder's value of $\delta_\mathrm{m}$, and the resulting PDFs are averaged for comparison with the count histograms.

The simplest model is a Poisson distribution whose expectation value incorporates linear bias only, and in the first column of Fig.~\ref{fig:pdfs}, the histogram of counts in cells for the low-$\alpha_0$ HOD in the first row (\mbox{$\alpha_0=0.798$}, \mbox{$\alpha_1=0.221$}) has a smaller variance than the overplotted Poisson curve, while the counts in cells for the high-$\alpha_0$ HOD (second row, \mbox{$\alpha_0=1.327$}, \mbox{$\alpha_1=0.772$}) have super-Poisson scatter at \mbox{$\delta_\mathrm{m}=0$} (red histogram). Not only does the Poisson model fail to match the variance of the measured distributions, it also does not track the mean accurately as $\delta_\mathrm{m}$ changes, suggesting that linear bias alone is insufficient to model the expectation value of the PDF for these HODs. The model with linear bias and stochasticity successfully matches the shape of the distribution, as it decouples the variance from the expectation value, but suffers from an offset to higher $N_\mathrm{g}$ for both nonzero values of $\delta_\mathrm{m}$. Finally, when quadratic bias is modeled in the third column of Fig.~\ref{fig:pdfs}, both the width and mean of the distribution accurately track the measured PDFs.

\subsection{Halo occupation distributions}
\label{subsec:hods}

For a given galaxy sample, assessing the range of plausible $\alpha_0$ and $\alpha_1$ values requires a means of pushing the high-dimensional space of galaxy-matter connections to various limits while still respecting constraints imposed by a given cosmological model and the galaxy catalog. Halo occupation distributions (HODs) are an efficient, empirically-motivated, and thoroughly studied approach to modeling this connection and generating mock galaxy catalogs within halos identified in $N$-body simulations \citep{Peacock_Smith,Berlind,Zheng_2005,Zheng_2007}. HOD modeling entails the assumptions that galaxies only exist inside dark matter halos and that the properties of a halo determine the expected number of galaxies it hosts. In its most commonly used form, an HOD is a function of halo mass which returns the expectation value of galaxy count for a halo. As such, HODs constitute both a modeling framework for cosmological analyses and a numerical method for painting galaxies into halos in dark matter-only simulations.

HODs are a natural choice for simulating and understanding galaxy stochasticity given their flexibility in describing a variety of galaxy distributions and the physical intepretability of their parameters. One of the most well-studied of these models is the 5-parameter HOD of \citet{Zheng_2005, Zheng_2007}, which expresses the expected counts of central and satellite galaxies as sigmoid and power law functions of halo mass, respectively. Here we consider a modified Zheng HOD of the form
\begin{align}
    \label{eq:zheng_cen}
    \Big\langle N_\mathrm{cen}\big(M_\mathrm{h}\big)\Big\rangle\,&=\,\frac{f_\mathrm{cen}}{2} \left[\,1+\mathrm{erf}\left( \frac{\log_{10} M_\mathrm{h}-\log_{10}M_\mathrm{min}}{\sigma_{\log\hspace{-0.1em}M}}\right)\,\right]\\[0.5em]
    \label{eq:zheng_sat}
    \Big\langle N_\mathrm{sat}\big(M_\mathrm{h}\big)\Big\rangle\,&=\,\Big\langle N_\mathrm{cen}\big(M_\mathrm{h}\big)\Big\rangle\,\times\,\left[\frac{M_\mathrm{h}}{M_1}\right]^{\,\alpha_{\,\mathrm{hod}}},
\end{align}
where masses are in units of $h^{-1}M_{\odot}$ and erf is the error function:
\begin{equation}
    \mathrm{erf}(x)\,\equiv\,\frac{2}{\sqrt{\pi}}\int_{\,0}^{\,x}e^{-t^{\,2}}dt\,.
\end{equation}
Relative to the canonical Zheng HOD, this version omits the cutoff mass $M_0$ in the numerator of $\langle N_\mathrm{sat}\rangle$ \citep[as in e.g.][]{Clampitt,Zacharegkas} and includes $f_\mathrm{cen}$ as a prefactor in $\langle N_\mathrm{cen}\rangle$, keeping the total at 5 model parameters.

For central galaxies, $M_\mathrm{min}$ sets the halo mass for which the probability of hosting a central is 0.5, and $f_\mathrm{cen}$ is an incompleteness parameter introduced to account for the failure of some fraction of those central galaxies to pass catalog selection, even in very massive halos \citep{Clampitt,Rodriguez_fcen,Leauthaud_fcen,Guo_2018,Zacharegkas}. The parameter $\sigma_{\log\hspace{-0.1em}M}$ describes the width of the mass range about $M_\mathrm{min}$ over which $\langle N_\mathrm{cen}\rangle$ transitions from zero to $f_\mathrm{cen}/2$.

For satellite galaxies, $M_1$ plays a role analogous to that of $M_\mathrm{min}$, characterizing the halo mass above which satellites are expected to form, and $\alpha_\mathrm{hod}$ is the power-law slope describing the increase in satellite count with increasing halo mass. The subscript on $\alpha_\mathrm{hod}$ is used throughout this work to avoid confusion with the stochasticity parameters $\alpha_0$ and $\alpha_1$. Note that in our chosen form of the Zheng HOD, $\langle N_\mathrm{sat}\rangle$ is modulated by $\langle N_\mathrm{cen}\rangle$, suppressing the placement of satellites in halos which have lower probabilities of hosting a central. As a result, $\langle N_\mathrm{sat}\rangle$ includes $f_\mathrm{cen}$ as a prefactor, albeit one fully degenerate with $M_1$. This HOD parametrization is therefore functionally equivalent to one in which $f_\mathrm{cen}$ does not appear in the satellite component, and one must simply account for the presence or absence of $f_\mathrm{cen}$ in $\langle N_\mathrm{sat}\rangle$ when comparing values of $M_1$ across models \citep[e.g.][]{Clampitt,Zacharegkas}.

An HOD in the form of equations \ref{eq:zheng_cen} and \ref{eq:zheng_sat} is useful both as a model for describing data and as a numerical method for placing mock galaxies in simulated halos. To generate a mock catalog using an HOD, an integer number of galaxies is placed in each halo, typically by drawing the central count \mbox{(0 or 1)} from a Bernoulli distribution with mean $\langle N_\mathrm{cen}\rangle$ and the satellite count from a Poisson distribution with expectation value $\langle N_\mathrm{sat}\rangle$. Central galaxies are placed at the centers of their host halos, and various approaches exist for positioning the satellites, including sampling from an analytical density profile such as NFW \citep{NFW} or assigning satellites to the positions of dark matter particles sampled from the halo (see e.g. \citealt{AbacusHOD} and Sect.~\ref{subsec:abacus}).

\subsection{HOD extensions: assembly bias}
The Zheng HOD and variants such as equations \ref{eq:zheng_cen} and \ref{eq:zheng_sat} have proven sufficiently flexible for modeling the moderate- to large-scale two-point correlation functions of LRG samples in current galaxy surveys such as DES \citep{Clampitt,Zacharegkas}. However, it has long been understood that secondary properties such as halo concentration can impact occupation statistics \citep{Wechsler_2002,Wechsler_2006,Mao_2017,Salcedo_2018,Hadzhiyska_2020,Xu_2021}. The dependence of galaxy count on any secondary (i.e. non-mass) halo properties is collectively referred to as galaxy assembly bias.

In this work we consider two proxies for assembly bias: halo concentration $c$ and the mean matter overdensity $\delta_\mathrm{env}$ of a halo's local environment. Concentration is a measure of the radial mass distribution within a halo, typically defined as the ratio of two characteristic radii, and is indicative of a halo's formation history. In this work we calculate concentration as implemented in \texttt{AbacusHOD} \citep[][see also Sect.~\ref{subsec:abacus} below]{AbacusHOD}, which defines it as \mbox{$c\equiv r_{90}/r_{25}$}, i.e. the ratio of the radii that contain 90\% and 25\% of the halo's total mass, respectively. Environmental density is determined by the total mass of all nearby halos whose centers lie beyond $r_{98}$ for the halo in question but within some outer radius, which we set at \mbox{$5\,h^{-1}\mathrm{Mpc}$} \citep[see][]{Xu_2021,AbacusHOD}. To define an overdensity, this mass is normalized by the mean of all such environmental masses around all halos:
\begin{equation}
    \delta_{\mathrm{env}}\,\equiv\,\frac{M_{\mathrm{h}}\big(r_{98}<r<5\,h^{-1}\mathrm{Mpc}\big)}{\big\langle M_{\mathrm{h}}\big(r_{98}<r<5\,h^{-1}\mathrm{Mpc}\big)\big\rangle}\,-\,1\,.
\end{equation}
Dependence on secondary halo properties such as concentration and environmental overdensity can be added to an HOD in various ways. Again in keeping with the \texttt{AbacusHOD} implementation, we follow the approach of \citet{Xu_2021}, which preserves the explicit form of the HOD in equations \ref{eq:zheng_cen} and \ref{eq:zheng_sat} by mixing assembly bias into the values of $M_\mathrm{min}$ and $M_1$ (see \citealt{Walsh_Tinker} for a similar approach):
\begin{align}
    \label{eq:mmin_mod}
    &\log_{10}M_\mathrm{min}\notag\\
    &\qquad\rightarrow\ \log_{10}M_\mathrm{min}+A_\mathrm{cent}\,\big(c_\mathrm{rank}-0.5\big)+B_\mathrm{cent}\,\big(\delta_\mathrm{rank}-0.5\big)\\[0.5em]
    \label{eq:m1_mod}
    &\log_{10}M_1\notag\\
    &\qquad\rightarrow\ \log_{10}M_1+A_\mathrm{sat}\,\big(c_\mathrm{rank}-0.5\big)+B_\mathrm{sat}\,\big(\delta_\mathrm{rank}-0.5\big)\ .
\end{align}
Here $A_\mathrm{cent}$, $A_\mathrm{sat}$, $B_\mathrm{cent}$, and $B_\mathrm{sat}$ are additional HOD parameters capturing assembly bias. Concentration and environmental overdensity are converted to ranked quantities $c_\mathrm{rank}$ and $\delta_\mathrm{rank}$, respectively, by binning halos by mass, ranking $c$ and $\delta_\mathrm{env}$ within each bin, and normalizing the ranks within bins to the range $[0,1]$. In an HOD with \mbox{$A_\mathrm{cent}>0$}, for example, a halo with above-median concentration for its mass bin (\mbox{$c_\mathrm{rank}>0.5$}) will see an increase in its effective value of $M_\mathrm{min}$ and thus a decrease in $\langle N_\mathrm{cen}\rangle$ relative to the case with no assembly bias.

In the context of studying galaxy stochasticity, the inclusion of assembly bias probes additional degrees of freedom in the galaxy-matter connection, in principle allowing the HOD to produce distributions $P(N_\mathrm{g\,}\vert\,\delta_\mathrm{m})$ which are physically plausible but cannot be realized when halo occupation depends on halo mass alone.


\section{Methods}
\label{sec:methods}

Our approach to studying galaxy stochasticity, described in detail in this section, can be summarized as follows: we first optimize the $M_\mathrm{min}$ and $M_1$ parameters of the HOD to reproduce the galaxy bias and density of a \textsc{redMaGiC} sample \citep{Rozo_redmagic}, at fixed fiducial values of the other HOD parameters. We then allow one other HOD parameter at a time to vary from its baseline value until $M_\mathrm{min}$ and $M_1$ can no longer be optimized to achieve the desired bias and density. Having derived a set of \textsc{redMaGiC}-like HODs in this way, we use them to produce mock galaxy catalogs and measure the resulting stochasticity. This approach allows us to determine relationships between individual HOD parameters and stochasticity at fixed bias and density, as well as to probe the allowed range of stochasticity when a limited number of HOD parameters are varied at once. We also assess the impact of cosmology on stochasticity by re-deriving a baseline HOD in each of a set of alternate cosmologies. Finally, to more thoroughly probe the full HOD parameter space, we Monte Carlo sample additional HODs with all parameters free and measure stochasticity in the resulting galaxy catalogs.

\subsection{Numerical simulations and HOD implementation}
\label{subsec:abacus}
We employ the \textsc{AbacusSummit}\footnote{\url{https://abacussummit.readthedocs.io}} suite of $N$-body simulations \citep{maksimova,Garrison2021} and associated halo catalogs produced by the \textsc{CompaSO} halo finder \citep{compaSO}, which are together designed to exceed DESI cosmological simulation requirements. We work with snapshots at \mbox{$z=0.3$} throughout; for reference, the density split analysis of \citet{Gruen_2018} used a \textsc{redMaGiC} LRG sample in the range \mbox{$0.2<z<0.45$}. The snapshot halo catalogs used in this work have been `cleaned' in a post-processing step described in \citet{Bose2022} by using merger trees to re-associate halos which are overdeblended by the halo finder. The \texttt{base} simulations of \textsc{AbacusSummit} consist of $6912^{3}$ particles in boxes of side length $2\,h^{-1}\mathrm{Gpc}$, with a particle mass of approximately $2\times10^{\,9}\,h^{-1}M_\odot$. Halos in the \texttt{base} simulations are resolved down to masses of roughly $10^{11}\,h^{-1}M_\odot$, which is more than sufficient for use with the HODs derived in this work. To reduce computing time, we apply a halo mass cut at \mbox{$\log_{10}(M_\mathrm{h}/h^{-1}M_\odot)=11.35$}, which is conservative given that we additionally constrain our HODs (see Sect.~\ref{subsubsec:additional_constraints}) such that that none will place a meaningful fraction of galaxies in halos below \mbox{$\log_{10}(M_\mathrm{h}/h^{-1}M_\odot)=11.5$}.

Our baseline analyses are performed in the primary cosmology (\texttt{c000}) of \textsc{AbacusSummit}, corresponding to Planck 2018 $\Lambda$CDM \citep[][\mbox{$\Omega_\mathrm{cdm}=0.2645$}, \mbox{$\Omega_\mathrm{b}=0.0493$}, \mbox{$n_\mathrm{s}=0.9649$}, \mbox{$\sigma_{8,\mathrm{m}}=0.8080$}, \mbox{$h=0.6736$}, \mbox{$N_\mathrm{eff}=3.046$}]{Planck2018}. For assessing the sensitivity of stochasticity to cosmology, we use the 52 additional cosmologies of the \textsc{AbacusSummit} emulator grid (\texttt{c130-181}; see \citealt{maksimova} section 2.2 for details on the grid selection).

For all HOD implementations, we use the \texttt{AbacusHOD} code \citep{AbacusHOD}, which is part of the \texttt{abacusutils}\footnote{\url{https://abacusutils.readthedocs.io}} package developed for working with \textsc{AbacusSummit} data products. \texttt{AbacusHOD} is a framework for rapidly generating mock galaxy samples in large simulation volumes and includes the \citet{Zheng_2007} model and assembly bias extensions \citep{Yuan_hod_ext} described in Sect.~\ref{subsec:hods} (see \citealt{Yuan_2021} for a summary of these and other HOD extensions). In this implementation, central galaxies are placed at halo centers (see \citealt{compaSO} section 2.2.2 for details on center identification in \textsc{CompaSO}). Each satellite galaxy is placed at the position of a different dark matter particle drawn at random from the halo. All galaxy catalogs in this work implement redshift space distortions (RSDs) by adjusting galaxy positions in the line-of-sight direction (i.e. parallel to the cylinders used to compute counts in cells) according to the velocity of the halo center or particle to which the galaxy was assigned. In the context of this work, these shifts in position model the expected effect of peculiar velocities on counts-in-cells statistics in a galaxy survey, where counts are calculated using galaxies within a redshift bin. A galaxy whose comoving distance corresponds to a Hubble-flow redshift within the bin may be excluded if its peculiar velocity leads to an observed redshift outside the bin, and vice versa. For completeness, we also measure the impact of turning off RSD modeling and find a sub-percent mean shift in galaxy count per cylinder and a similar sub-percent effect on linear bias. The effect on the stochasticity parameters $\alpha_0$ and $\alpha_1$ is at the few percent level, which is subdominant to changes in stochasticity stemming from degrees of freedom in the galaxy-matter connection, as probed by the range of HODs we derive here.

\subsection{Selecting plausible HODs}
\label{subsec:opt_hods}
\subsubsection{Constraining galaxy bias and density}
\label{subsubsec:constrain_b_n}
Because we aim to study the stochasticity of LRGs, matching known properties of a \textsc{redMaGiC} galaxy sample is a natural choice. \citet{Gruen_2018} applied the stochasticity model of equation~\ref{eq:stoch} to a \textsc{redMaGiC} high-density sample, which by construction has a mean comoving density \mbox{$n=10^{-3}\,h^3\,\mathrm{Mpc}^{-3}$}, and we set this as our target value of galaxy density. We simultaneously optimize our HODs for a baseline linear galaxy bias of \mbox{$b=1.5$} in the model of equation~\ref{eq:stoch}, similar to the \mbox{$b=1.54$} found by \citet{Friedrich_2018} when applying the same model to mock LRGs across a range of smoothing scales from 10 to 30 arcmin. Although we select \textsc{redMaGiC}-like values here, we emphasize that $b$ and $n$ in this method may be set to values corresponding to any desired galaxy population. To assess the impact of bias and density on stochasticity, we construct two additional sets of HODs which vary the target values of $b$ and $n$ themselves (see Sect.~\ref{subsubsec:hod_curves}).

Optimizing for a specific bias and density requires a means of calculating these properties at different points in HOD parameter space. For an HOD which depends on halo mass only (i.e. no assembly bias), linear galaxy bias and number density may be estimated efficiently using analytic forms for the halo mass function $dn_\mathrm{h}/dM_\mathrm{h}$ \citep[e.g.][]{Tinker_2010} and halo bias $b_\mathrm{h}$ as a function of mass \citep{Tinker_2008}:
\begin{align}
    n\,&=\,\int\frac{dn_\mathrm{h}}{dM_\mathrm{h}}\,\Big\langle N_\mathrm{g}\big(M_\mathrm{h}\big)\Big\rangle\,dM_\mathrm{h}\\[0.5em]
    b\,&=\,\frac{1}{n}\int\frac{dn_\mathrm{h}}{dM_\mathrm{h}}\,\Big\langle N_\mathrm{g}\big(M_\mathrm{h}\big)\Big\rangle\,b_\mathrm{h}\big(M_\mathrm{h}\big)\,dM_\mathrm{h}\,,
\end{align}
where $\langle N_\mathrm{g}(M_\mathrm{h})\rangle$ is the expected galaxy count given by the HOD. However, we wish to construct HODs which include assembly bias, in which case analytic estimates analogous to those above would require integration over halo concentration and environmental overdensity -- and therefore also approximations for $dn_\mathrm{h}/dM_\mathrm{h}$ and $b_\mathrm{h}$ which include dependence on these secondary halo properties. Lacking such functions, we instead estimate the bias and density of an HOD directly by using \texttt{AbacusHOD} to generate a single mock galaxy catalog. Calculating the number density of the catalog is straightforward given the simulation volume and the total galaxy count. To measure galaxy bias while optimizing HODs, we follow the approach of \citet{Friedrich_2018} and \citet{Gruen_2018} and fit the linear bias and stochasticity model of equation~\ref{eq:stoch} to galaxy counts and matter overdensities in cells (see Sect.~\ref{subsubsec:cics_stoch} for details on model fitting).

Given the above method for measuring linear bias and density, the optimization procedure begins by setting all parameters except $M_\mathrm{min}$ and $M_1$ to chosen baseline values (Sect.~\ref{subsubsec:fiducial_hod}). The HOD then has two free parameters and can be optimized for the unique combination $(M_\mathrm{min},M_1)$ that satisfies the two constraints (bias and density), under the assumption that a unique optimal HOD exists within a physically plausible range of $(M_\mathrm{min},M_1)$.\footnote{We have checked that this is indeed a reasonable assumption by using fitting functions for halo bias \citep{Tinker_2010} and the halo mass function \citep{Tinker_2008}, together with our HODs, to analytically model galaxy bias and density as functions of $M_\mathrm{min}$ and $M_1$, at various combinations of the other HOD parameters.} Both mass parameters are bounded between $10^{11}$ and $10^{16}\,h^{-1}M_\odot$, fully capturing the range of halo masses in the baseline cosmology of \textsc{AbacusSummit} (subject to our mass cut). The optimization itself consists of minimizing (over $M_\mathrm{min}$ and $M_1$) the sum of squared relative errors in bias and density given the goals of \mbox{$b=1.5$} and \mbox{$n=10^{-3}\,h^3\,\mathrm{Mpc}^{-3}$}, i.e. minimizing the function
\begin{equation}
    f(b,n)\ \equiv\ \left(\frac{b}{1.5}-1\right)^2+\,\Bigg(\frac{n}{10^{-3}\,h^3\,\mathrm{Mpc}^{-3}}-1\,\Bigg)^2
\end{equation}
via Nelder-Mead simplex \citep{Nelder-Mead}. Although measuring bias and density directly on a mock catalog is naturally more computationally expensive than an analytic estimate, optimizing an HOD to sub-percent error in both $b$ and $n$ is generally possible in $\sim$50 function evaluations.

\subsubsection{Fiducial HOD parameters}
\label{subsubsec:fiducial_hod}

The optimization procedure described in Sect.~\ref{subsubsec:constrain_b_n} requires a specification of all HOD parameters other than $M_\mathrm{min}$ and $M_1$. In this section, we discuss the choice of fiducial values for these non-mass parameters. Optimizing an HOD at this set of values yields a baseline model relative to which other HODs can be derived.

Our fiducial value of $f_\mathrm{cen}$ is a direct estimate from the DES \texttt{Y3\;GOLD} \textsc{redMaGiC} high-density catalog, selecting only galaxies in the same redshift range (\mbox{$0.2<z<0.45$}) used by \citet{Gruen_2018}. The completeness of central galaxies in the sample is estimated by summing the probabilities that each of the selected \textsc{redMaGiC} galaxies is a central in a \textsc{redMaPPer} cluster, based on candidate centrals identified by \textsc{redMaPPer} \citep{redmapper_2014,redmapper_2016} in the same \texttt{Y3\;GOLD} photometric dataset \citep{Sevilla-Noarbe_2021}. We select clusters of richness \mbox{$\lambda\geq50$}, corresponding to halos with masses well above $M_{\mathrm{min}}$ (see e.g. figure 9 in \citealt{DESY1_clusters}), so that \mbox{$\langle N_\mathrm{cen}\rangle\simeq f_\mathrm{cen}$} for these clusters. The sum of these central probabilities divided by the number of clusters gives an estimate of \mbox{$f_\mathrm{cen}=0.316$}, which we take as our fiducial value.

We set \mbox{$\sigma_{\log\hspace{-0.1em}M}=0.3$} as a baseline value for the width of the $\langle N_\mathrm{cen}\rangle$ part of the HOD, consistent with published constraints for \textsc{redMaGiC} samples for lens bins of a similar redshift range \citep{Clampitt,Zacharegkas}. We find that stochasticity as parametrized by $\alpha_0$ and $\alpha_1$ is less sensitive to the choice of $\sigma_{\log\hspace{-0.1em}M}$ than to the other HOD parameters tested. 

For the satellite galaxy component of the HOD, we set the fiducial power-law slope to \mbox{$\alpha_\mathrm{hod}=1$}. This value is a significant reduction relative to the DES Year 3 \textsc{redMaGiC} HOD of \citet{Zacharegkas} for the lowest-redshift lens bin. We find that slopes much larger than \mbox{$\alpha_\mathrm{hod}=1$} yield galaxy counts in the most massive halos of the baseline \textsc{AbacusSummit} simulation which are a factor of several larger than the highest \textsc{redMaGiC} counts in clusters (see Sect.~\ref{subsubsec:constrain_ngal}). A similar effect was noted by \citet{Kokron_2022}, who found that a reduction in $\alpha_\mathrm{hod}$ relative to the DES Y3 values was needed to match derived parameters such as the satellite fraction of their \textsc{redMaGiC}-like HOD. The authors attributed the discrepancy in this case to differences between the halo mass function of their simulations and the \citet{Tinker_2008} mass function used in constraining the DES Y3 HOD \cite[see section 3 of][]{Kokron_2022}.

Finally, we define our baseline model without assembly bias, setting $A_\mathrm{cent}$, $A_\mathrm{sat}$, $B_\mathrm{cent}$, and $B_\mathrm{sat}$ to zero. Table~\ref{tab:baseline_hod} lists the parameters of the baseline HOD after optimizing for $M_\mathrm{min}$ and $M_1$. To determine the extent to which each HOD parameter may vary in isolation, we choose one non-mass parameter at a time and shift it to values above and below its baseline, re-optimizing $M_\mathrm{min}$ and $M_1$ each time to obtain an HOD which conserves the target bias and density. For sufficiently extreme values of each parameter, the constraints can no longer be satisfied within the bounds set on $M_\mathrm{min}$ and $M_1$ (from $10^{11}$ to $10^{16}\,h^{-1}M_\odot$ for both mass parameters). The same applies when varying the target value of bias or density with all non-mass HOD parameters fixed at their baseline values. We use these ranges for the individual parameters (Table~\ref{tab:baseline_hod}) to derive additional HODs in Sect.~\ref{subsubsec:hod_curves}.

\begin{table}
    \caption{Parameter values for the baseline HOD and ranges for the HOD curves described in Sect.~\ref{subsubsec:hod_curves}. Note that the values of $M_\mathrm{min}$ and $M_1$ for any derived HOD (including the baseline) are the result of the optimization procedure described in Sect.~\ref{subsec:opt_hods} and therefore do not have a single fixed minimum and maximum across the various HOD curves. The two mass parameters are given in units of $h^{-1}M_\odot$, and densities are in units of $h^3\,\mathrm{Mpc}^{-3}$.}
    \label{tab:baseline_hod}
    \begin{center}
        \begin{tabular}{lccc}
            Parameter & Baseline & Min & Max\\
            \noalign{\smallskip}
            \hline
            \hline
            \noalign{\smallskip}
            $\log_{10}M_\mathrm{min}$ & 12.32 & -- & --\\[0.4em]
            $\log_{10}M_1$ & 13.26 & -- & --\\[0.4em]
            $\sigma_{\log\hspace{-0.1em}M}$ & 0.3 & 0.01 & 0.38\\[0.4em]
            $\alpha_\mathrm{hod}$ & 1.0 & 0.85 & 1.03\\[0.4em]
            $f_\mathrm{cen}$ & 0.316 & 0.25 & 0.60\\[0.4em]
            $A_\mathrm{cent}$ & 0 & $-\,$0.40\phantom{...} & 0.44\\[0.4em]
            $A_\mathrm{sat}$ & 0 & $-\,$0.55\phantom{...} & 0.56\\[0.4em]
            $B_\mathrm{cent}$ & 0 & $-\,$0.20\phantom{...} & 0.09\\[0.4em]
            $B_\mathrm{sat}$ & 0 & $-\,$1.00\phantom{...} & 0.23\\[0.4em]
            \hline
            \noalign{\smallskip}
            bias & 1.5 & 1.36 & 1.54\\[0.4em]
            density & $1.0\times10^{-3}$ & $7.2\times10^{-4}$ & $1.08\times10^{-3}$\\
            \noalign{\smallskip}
            \hline
        \end{tabular}
    \end{center}
\end{table}

\subsubsection{Constraining maximum halo occupation}
\label{subsubsec:constrain_ngal}

\begin{figure}
    \centering
	\includegraphics[width=\columnwidth]{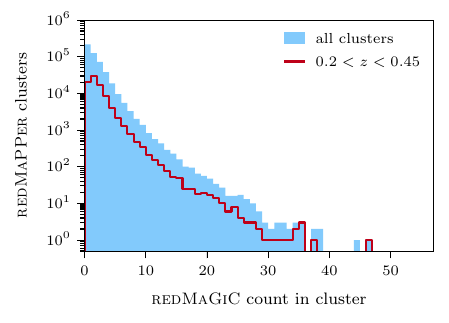}
    \caption{The distribution of counts of DES \texttt{Y3\;GOLD} \textsc{redMaGiC} galaxies (high-density sample) in \textsc{redMaPPer} clusters. Count per cluster is calculated as the sum of membership probabilities for all \textsc{redMaGiC} galaxies identified as possible members of the cluster. Shown are the distributions for all clusters in the catalog (blue) and for clusters in the redshift range \mbox{$0.2<z<0.45$} (red); this is identical to the range used in the density split analysis of DES Y1 \textsc{redMaGiC} galaxies in \protect\citet{Gruen_2018}.}
    \label{fig:redmagic}
\end{figure}

With galaxy bias and density as constraints, the optimization procedure remains free to derive HODs whose mock galaxy catalogs may differ from those of a \textsc{redMaGiC} sample in properties other than bias and density. In particular, we find that it is possible to obtain HODs which satisfy \mbox{$b\simeq1.5$} and \mbox{$n\simeq10^{-3}\,h^3\,\mathrm{Mpc}^{-3}$} but whose slopes $\alpha_\mathrm{hod}$ are high enough to yield \mbox{$\langle N_\mathrm{sat}\rangle\gtrsim100$} for the most occupied halos. For comparison to observations, we estimate the maximum number of galaxies in a \textsc{redMaGiC} catalog (of the same density) which appear in any one galaxy cluster. For this we again use the DES \texttt{Y3\;GOLD} \textsc{redMaGiC} catalog (high-density, \mbox{$n=10^{-3}\,h^3\,\mathrm{Mpc}^{-3}$}) and the \texttt{Y3\;GOLD} \textsc{redMaPPer} cluster catalog. To estimate the count of \textsc{redMaGiC} galaxies in a given cluster, we sum the \textsc{redMaPPer} membership probabilities for \textsc{redMaGiC} galaxies identified as potential members of that cluster. The results are shown in Fig.~\ref{fig:redmagic}, which includes the distribution of cluster occupations for all \textsc{redMaPPer} clusters and for the subset in the redshift range \mbox{$0.2<z<0.45$}, the same as in the density split analysis of DES Y1 \textsc{redMaGiC} galaxies \citep{Gruen_2018}.

To keep our selected HODs consistent with the high end of the distribution in Fig.~\ref{fig:redmagic}, we set an additional constraint on the single largest value of \mbox{$\langle N_\mathrm{g}\rangle=\langle N_\mathrm{cen}\rangle+\langle N_\mathrm{sat}\rangle$} across all halos, requiring that it fall in the range $[25,50]$. Note that the constraint is on the expectation value $\langle N_\mathrm{g}\rangle$ and that any individual realization of a mock catalog may place more than 50 or fewer than 25 galaxies in the halo with the highest galaxy count, due to the random drawing of satellite galaxy counts. We find that this constraint on $\langle N_\mathrm{g}\rangle$ further tightens the bounds on the allowed values of the 7 non-mass HOD parameters relative to those imposed by bias and density alone. Placing limits on $\langle N_\mathrm{g}\rangle$ thus sets the extent to which each HOD parameter may vary in isolation as listed in Table~\ref{tab:baseline_hod} (with exceptions regarding the upper limit on $\sigma_{\log\hspace{-0.1em}M}$ and lower limit on $B_\mathrm{sat}$, see Sect.~\ref{subsubsec:additional_constraints}).

With this restriction on the maximum value of $\langle N_\mathrm{g}\rangle$ included, the full set of baseline constraints used in deriving HODs is:
\begin{align}
    \label{eq:constraints}
    &b\,=\,1.5\notag\\[0.4em]
    &n\,=\,10^{-3}\,h^3\,\mathrm{Mpc}^{-3}\notag\\[0.4em]
    &\max_i \Big\langle N_\mathrm{g}\big(M_{\mathrm{h},i}\big)\Big\rangle\,\in\,[25,50]\ ,
\end{align}
where the first two constraints are imposed during optimization for the mass parameters $(M_\mathrm{min},M_1)$ and the third limits the extent to which each non-mass parameter in isolation may deviate from its fiducial value.

\subsubsection{Additional constraints: \texorpdfstring{$\sigma_{\log\hspace{-0.1em}M}$}{sigma} and assembly bias}
\label{subsubsec:additional_constraints}

For the one set of HODs that varies $\sigma_{\log\hspace{-0.1em}M}$, we apply a further constraint due to the fact that if this parameter is sufficiently large, it will attempt to place galaxies in halos below our mass cut or even below the minimum mass resolved by the halo finder, regardless of the value of $M_\mathrm{min}$. We therefore set an upper limit on $\sigma_{\log\hspace{-0.1em}M}$ such that no HOD places a substantial fraction of galaxies (\mbox{$\gtrsim$$10^{-4}$}) in halos below \mbox{$\log_{10}(M_\mathrm{h}/h^{-1}M_\odot)=11.5$}. This condition corresponds to an upper bound of \mbox{$\sigma_{\log\hspace{-0.1em}M}\simeq0.38$}. Recall (Sect.~\ref{subsec:abacus}) that we use a halo catalog that is complete down to \mbox{$\log_{10}(M_\mathrm{h}/h^{-1}M_\odot)=11.35$}. Beyond being practical, this should also be a physically reasonable constraint on $\sigma_{\log\hspace{-0.1em}M}$, as we do not expect halos below \mbox{$\log_{10}(M_\mathrm{h}/h^{-1}M_\odot)=11.5$} to host a meaningful number of galaxies that pass the luminosity threshold and other selection criteria for a \textsc{redMaGiC} high-density sample, based on stellar mass to halo mass ratio considerations. We additionally verify that even without this constraint on $\sigma_{\log\hspace{-0.1em}M}$, and using a lower halo mass cut at \mbox{$\log_{10}(M_\mathrm{h}/h^{-1}M_{\odot})=11$}, the highest-$\sigma_{\log\hspace{-0.1em}M}$ HODs consistent with all other constraints (equation \ref{eq:constraints}) still produce values of $\alpha_0$ and $\alpha_1$ within the range spanned by the HODs that vary other parameters.

Finally, we restrict each of the four assembly bias parameters $A_\mathrm{cent}$, $A_\mathrm{sat}$, $B_\mathrm{cent}$, and $B_\mathrm{sat}$ to the range $[-1,1]$, which is generously wide relative to assembly bias values measured in observations and simulations \citep[e.g.][]{Xu_2021,AbacusHOD,Yuan_2021,Yuan2022,Hadzhiyska2023,Paviot2024}. In practice, we find that it is only necessary to impose this restriction at \mbox{$B_\mathrm{sat}=-1$}; all other assembly bias limits are set more tightly by the constraints in equation~\ref{eq:constraints} as long as all other HOD parameters are kept at their baseline values.

\subsubsection{HOD curves of constant bias and density}
\label{subsubsec:hod_curves}

For each of the 7 HOD parameters other than $M_\mathrm{min}$ and $M_1$, including assembly bias, we optimize HODs at 11 different values of that parameter, linearly spaced across its allowed range (see Sect.~\ref{subsubsec:fiducial_hod} and Table~\ref{tab:baseline_hod}), with the other 6 non-mass parameters set to their baseline values. Each optimization determines the values of $M_\mathrm{min}$ and $M_1$ that produce the target galaxy bias and density. We execute the same procedure across the allowed ranges of the target values of bias and density themselves, with the HOD (excluding $M_\mathrm{min}$ and $M_1$) fixed at its baseline parameters. The result is a set of 9 curves in HOD parameter space, all intersecting at the point corresponding to the baseline model. All HODs on the curves which vary the non-mass HOD parameters produce mock galaxy catalogs which conserve linear bias \mbox{$b=1.5$} (in the model of equation~\ref{eq:stoch}) and number density \mbox{$n=10^{-3}\,h^{3}\,\mathrm{Mpc}^{-3}$}, while the curves which vary bias or density have changing values of their respective parameter by construction.

\subsubsection{Counts in cells and measuring stochasticity}
\label{subsubsec:cics_stoch}

To make stochasticity measurements on mock catalogs generated by our HODs, we count galaxies in cylindrical cells of radius $10\,h^{-1}\mathrm{Mpc}$ (as a transverse comoving distance, equivalent to $\sim$30 arcmin at \mbox{$z=0.3$}) and depth $500\,h^{-1}\mathrm{Mpc}$ (roughly the comoving depth of a redshift bin between, e.g., \mbox{$z=0.23$} and \mbox{$z=0.36$}; note of course that each mock catalog exists in a snapshot at \mbox{$z=0.3$}). Each $2\,h^{-1}\mathrm{Gpc}$ simulation box is treated as 4 slabs of depth $500\,h^{-1}\mathrm{Mpc}$, and counts in cells are performed in all slabs in order to utilize the full simulation volume. Across the $(2\,h^{-1}\mathrm{Gpc})^2$ face of each slab, cylinders are placed in a \mbox{$200\times200$} grid with a $10\,h^{-1}\mathrm{Mpc}$ spacing between their centers, and periodic boundary conditions are enforced. Each mock catalog therefore yields a set of 160,000 counts in cells. Measurements of matter overdensity $\delta_\mathrm{m}$ in cells are performed similarly by counting dark matter particles from the \texttt{A} subsample provided by \textsc{AbacusSummit}, which uniformly samples 3 percent of the total particles in the simulation. A downsampling of particles naturally contributes some variance to the values of $\delta_\mathrm{m}$ in cells; however, even the most underdense cylinder (with radius $10\,h^{-1}\mathrm{Mpc}$ and depth $500\,h^{-1}\mathrm{Mpc}$ in the baseline cosmology) samples \mbox{$>$$10^5$} particles, and so any matter variance is subdominant to the stochasticity contributed by the galaxy-matter connections considered here.

To fit the stochasticity model to the resulting set of tuples $(N_{\mathrm{g},i},\delta_{\mathrm{m},i})$, we take our model likelihood to be the product over all cells of the probability mass function of equation~\ref{eq:stoch_quad}. As noted in \citet[][sec. IV.C.2]{Friedrich_2018}, although this likelihood is not exact given the correlations between counts and densities in nearby cells, it is nonetheless sufficient for the purpose of this analysis, namely obtaining conservative bounds on $\alpha_0$ and $\alpha_1$. The logarithm of our likelihood is then the sum over all cells of the logarithm of the probability mass given by equation~\ref{eq:stoch_quad}:
\begin{align}
    \label{eq:lnlike}
    \ln\mathcal{L}\,=\sum_i\left[\rule{0em}{2em}\right.&-\ln\big(\alpha_0+\alpha_1\delta_{\mathrm{m},i}\big)-\frac{\bar{N}_\mathrm{g}\,\Big[1+b_1\delta_{\mathrm{m},i}+\frac{b_2}{2}\big(\delta_{\mathrm{m},i}^{\,2}-\sigma_\mathrm{m}^{\,2}\big)\Big]}{\alpha_0+\alpha_1\delta_{\mathrm{m},i}}\notag\\[0.5em]
    &+\,\frac{N_{\mathrm{g},i}}{\alpha_1+\alpha_1\delta_{\mathrm{m},i}}\ln\left(\frac{\bar{N}_{\mathrm{g}}\,\Big[1+b_1\delta_{\mathrm{m},i}+\frac{b_2}{2}\big(\delta_{\mathrm{m},i}^{\,2}-\sigma_\mathrm{m}^{\,2}\big)\Big]}{\alpha_0+\alpha_1\delta_{\mathrm{m},i}}\right)\notag\\[0.5em]
    &\hspace{6em}-\,\ln\Gamma\left(\frac{N_{\mathrm{g},i}}{\alpha_0+\alpha_1\delta_{\mathrm{m},i}}+1\right)\left.\rule{0em}{2em}\right]\,.
\end{align}
Note that this likelihood assumes the \mbox{$1/\alpha$} normalization for $P(N_\mathrm{g\,}\vert\,\delta_\mathrm{m})$, which we find to be accurate for all stochasticity values measured for the HODs in this work.

For a given parameter combination $b_1,b_2,\alpha_0,\alpha_1$, it may be the case that some cells have $\delta_{\mathrm{m},i}$ such that the quantities \mbox{$1+b\delta_{\mathrm{m},i}+\frac{b_2}{2}(\delta_{\mathrm{m},i}^{\,2}-\sigma_\mathrm{m}^{\,2})$} and/or \mbox{$\alpha_0+\alpha_1\delta_{\mathrm{m},i}$} are zero or negative. These cases may be handled in a number of ways, for instance by setting the offending quantity to a very small positive value or returning negative infinity for $\ln\mathcal{L}$. We use the latter option and verify that the two methods do not differ significantly in terms of the values of $b_1$, $b_2$, $\alpha_0$, and $\alpha_1$ that maximize the likelihood of equation~\ref{eq:lnlike}.


\section{Results}
\label{sec:results}

\subsection{Selected HODs}
\label{subsec:hod_results}

\begin{figure}
    \includegraphics[width=\columnwidth]{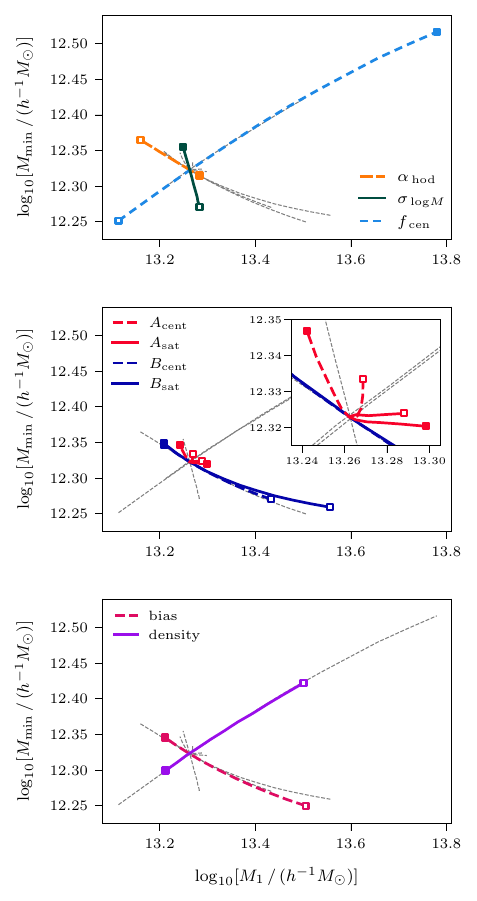}
    \caption{$M_\mathrm{min}$ versus $M_1$ for the HODs produced when one non-mass parameter is varied and $M_1$ and $M_\mathrm{min}$ are optimized to achieve the target bias and density. The ends of each curve are marked with an open square for the lowest value of that parameter and a filled square for the highest -- see Table~\ref{tab:baseline_hod} for the parameter ranges. All HOD curves intersect at the location of the baseline model. Top panel: HODs which vary the parameters $\alpha_\mathrm{hod}$, $\sigma_{\log\hspace{-0.1em}M}$, and $f_\mathrm{cen}$. Middle panel: HODs which vary the assembly bias parameters $A_\mathrm{cent}$, $A_\mathrm{sat}$, $B_\mathrm{cent}$, and $B_\mathrm{sat}$ (inset: closeup of the curves for $A_\mathrm{cent}$ and $A_\mathrm{sat}$). Bottom panel: HODs which vary the target values of bias and density while keeping the non-mass parameters fixed at baseline. Gray dashed curves in each panel correspond to the curves from the other two panels for reference.}
    \label{fig:hod_curves}
\end{figure}

The HOD curves for varying $\alpha_\mathrm{hod}$, $\sigma_{\log\hspace{-0.1em}M}$, and $f_\mathrm{cen}$ are shown in the $M_1$-$M_\mathrm{min}$ projection in HOD parameter space in the top panel of Fig.~\ref{fig:hod_curves}. Each curve consists of 11 HODs linearly spaced in the parameter being varied, with the endpoints occurring where it is no longer possible to meet the constraints on bias, number density, and the maximum value of $\langle N_\mathrm{g}\rangle$ simultaneously. The maximum and minimum values for the parameter varied in each each curve are listed in Table~\ref{tab:baseline_hod}. We verify that these HODs, when used to generate new mock galaxy catalogs with unique random seeds, reproduce the target values of bias and density to sub-percent precision. Note that the three curves in the top panel of Fig.~\ref{fig:hod_curves} extend in orthogonal directions in the full HOD parameter space, as each varies a parameter which is fixed for the other curves.

Analogous results for the assembly bias parameters $A_\mathrm{cent}$, $A_\mathrm{sat}$, $B_\mathrm{cent}$, and $B_\mathrm{sat}$ are shown in the middle panel of Fig.~\ref{fig:hod_curves}. Unlike the curves for all other HOD parameters, those for $A_\mathrm{cent}$ and $A_\mathrm{sat}$ (parameters which modulate the effect of halo concentration) are U-shaped, with the HODs corresponding to negative and positive parameter values extending in somewhat similar directions in the $M_1$-$M_\mathrm{min}$ projection. The curve for $B_\mathrm{sat}$ is limited at the low-$B_\mathrm{sat}$ end by our conservatively wide lower bound of $-1$ for the assembly bias parameters, as noted in Sect.~\ref{subsubsec:additional_constraints}.

The bottom panel of Fig.~\ref{fig:hod_curves} shows the HODs that result when the target bias and density values themselves are varied. For each curve, the target value of the other parameter remains fixed at its baseline value (\mbox{$b=1.5$} or \mbox{$n=10^{-3}\,h^3\,\mathrm{Mpc}^{-3}$}), and the endpoints occur where the constraint on the maximum value of $\langle N_\mathrm{g}\rangle$ (equation~\ref{eq:constraints}) can no longer be satisfied for more extreme values of the parameter varied in that curve. The resulting ranges are \mbox{$[1.36,1.54]$} for bias and \mbox{$[0.72,1.08]\times10^{-3}\,h^{3}\,\mathrm{Mpc}^{-3}$} for density. Within these limits imposed by our constraints, varying the bias and density of the desired galaxy sample produces values of $M_\mathrm{min}$ and $M_1$ which fall roughly within the range spanned by the curves that vary HOD parameters in the first two panels of Fig.~\ref{fig:hod_curves}. However, this does not imply that the same should hold for the stochasticity values $\alpha_0$ and $\alpha_1$ that result when these HODs are used to generate mock galaxy catalogs.

\subsection{Galaxy stochasticity}
\label{subsec:stoch_results}

\begin{figure}
    \includegraphics[width=\columnwidth]{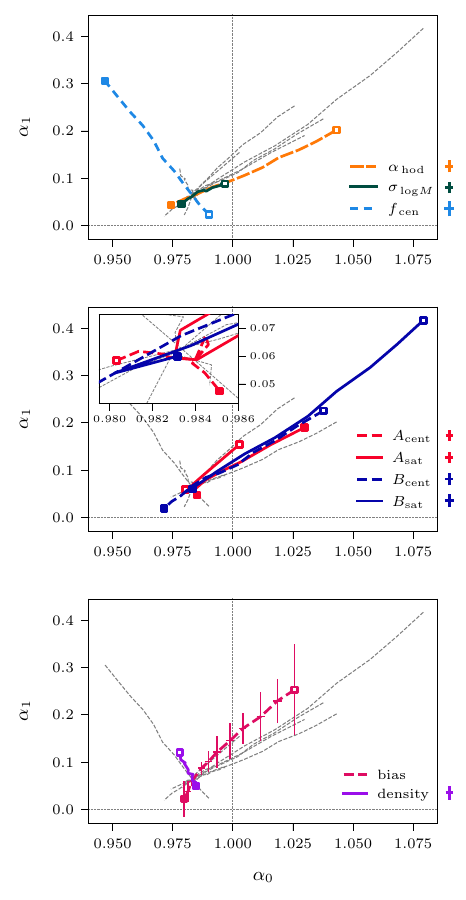}
    \caption{Stochasticity parameters $\alpha_0$ and $\alpha_1$ corresponding to the HOD curves of Fig.~\ref{fig:hod_curves}. Top panel: stochasticity for the curves that vary the HOD parameters $\alpha_\mathrm{hod}$, $\sigma_{\log\hspace{-0.1em}M}$, and $f_\mathrm{cen}$. Middle panel: stochasticity for the curves that vary the assembly bias parameters $A_\mathrm{cent}$, $A_\mathrm{sat}$, $B_\mathrm{cent}$, and $B_\mathrm{sat}$ (inset: closeup of the curve for $A_\mathrm{cent}$). Bottom panel: stochasticity for the curves that vary the target values of bias and density. Open and filled squares correspond to the HODs with the lowest and highest parameter values, respectively, for the parameter being varied in each curve. Error bars to the right of each panel indicate maximum spatial jackknife error in $\alpha_0$ and $\alpha_1$ across all 11 points in a curve. Errors are plotted point-by-point for galaxy bias (bottom panel) due to the larger range of error sizes for $\alpha_1$. Dashed lines at \mbox{$\alpha_0=1$} and \mbox{$\alpha_1=0$} indicate the stochasticity parameter values corresponding to Poisson shot noise.}
    \label{fig:stoch_curves}
\end{figure}

We evaluate stochasticity for the HODs of Fig.~\ref{fig:hod_curves} by generating 100 mock galaxy catalogs with unique random seeds for each of the 11 HODs in each curve. We compute counts in cylinders and measure bias and stochasticity ($b_1,b_2,\alpha_0,\alpha_1$) as described in Sect.~\ref{subsubsec:cics_stoch}, averaging the results across the 100 samples to mitigate the Bernoulli/Poisson noise in the drawing of galaxy counts. Uncertainty estimates are obtained by measuring stochasticity on spatial jackknife resamplings of the counts in cells data, removing square jackknife patches of \mbox{$20\times20$} cylinders at a time. Shown in the top panel of Fig.~\ref{fig:stoch_curves} are stochasticity curves corresponding to the HOD curves in the top panel of Fig.~\ref{fig:hod_curves} for the parameters $\alpha_\mathrm{hod}$, $\sigma_{\log\hspace{-0.1em}M}$, and $f_\mathrm{cen}$. All HODs with varying $\sigma_{\log\hspace{-0.1em}M}$ and $f_\mathrm{cen}$ have \mbox{$\alpha_0<1$}, corresponding to shot noise which is sub-Poisson for \mbox{$\delta_\mathrm{m}\leq0$} but super-Poisson for cells with sufficiently positive $\delta_\mathrm{m}$, given that \mbox{$\alpha_1>0$}. HODs with values of $\alpha_\mathrm{hod}$ near the lower end of the allowed range do produce \mbox{$\alpha_0>1$} and therefore super-Poisson shot noise even down to some negative matter overdensities. Among the three varying parameters shown in the top panel of Fig.~\ref{fig:stoch_curves}, none produce \mbox{$\alpha_1<0$}, and so shot noise in galaxy counts increases with local matter density in all 33 of these HODs. 

Analogous results for the four assembly bias parameters are shown in the middle panel of Fig.~\ref{fig:stoch_curves}. When varying $B_\mathrm{sat}$ in particular, we obtain larger values of both $\alpha_0$ and $\alpha_1$ than in all other curves, with the highest-stochasticity HOD being that with the largest $M_1$ and smallest $M_\mathrm{min}$ in the curve (the \mbox{$B_\mathrm{sat}=-1$} HOD in the lower right corner of the middle panel of Fig.~\ref{fig:hod_curves}). Recall that the assembly bias prescription of equations~\ref{eq:mmin_mod} and \ref{eq:m1_mod} modifies the effective values of $M_\mathrm{min}$ and $M_1$ on a per-halo basis depending on concentration and environmental density. In particular, \mbox{$B_\mathrm{sat}<0$} leads to reduced effective $M_1$ (and thus more satellites) in halos whose local environments are above median density for their halo mass bin. The result for \mbox{$B_\mathrm{sat}=-1$} therefore corresponds to stochasticity being largest (among these curves) when expected satellite galaxy count $\langle N_\mathrm{sat}\rangle$ is positively correlated with the density of a halo's local environment. For $A_\mathrm{cent}$ and $A_\mathrm{sat}$, the U-shaped nature of the curves in $M_1$-$M_\mathrm{min}$ space is reproduced in the plot of $\alpha_1$ versus $\alpha_0$. Due to the relatively small extent of the stochasticity results for $A_\mathrm{cent}$, sample variance in the finite set of mock galaxy catalogs has a strong effect on the smoothness of the $A_\mathrm{cent}$ curve in the middle panel of Fig.~\ref{fig:stoch_curves} (see inset).

Stochasticity results for the final two HOD curves, those that vary the target values of galaxy bias and density, are shown in the bottom panel of Fig.~\ref{fig:stoch_curves}. The results for varying density roughly share a degeneracy direction in the $\alpha_0$-$\alpha_1$ plane with those for varying $f_\mathrm{cen}$ (refer to the top panel of Fig.~\ref{fig:stoch_curves}). Recall that the range of target density for these HODs is \mbox{$[0.72,1.08]\times10^{-3}\,h^{3}\,\mathrm{Mpc}^{-3}$}. We find that increasing galaxy density (while the non-mass HOD parameters remain at their baseline values) leads to a slight increase in global stochasticity $\alpha_0$ and a decrease in $\alpha_1$. In contrast, increasing the target bias (over its allowed range from 1.36 to 1.54) drives down both stochasticity parameters. As with the results for varying HOD parameters, we find that all points on the curves of varying bias and density have \mbox{$\alpha_1>0$}, i.e. higher shot noise in galaxy counts at higher matter overdensity. Sufficiently low values of the target bias (again with the non-mass HOD parameters fixed at baseline) yield \mbox{$\alpha_0>1$}, corresponding to super-Poisson shot noise at \mbox{$\delta_\mathrm{m}=0$}. This observation further favors the use of highly biased tracers for cosmological studies.

\subsection{Stochasticity versus cosmology and cell geometry}
\label{subsec:stoch_results_cosmo}

To assess the extent to which uncertainty in the underlying cosmology affects galaxy stochasticity, we repeat the procedure for deriving the baseline HOD in each of 52 cosmologies from the \textsc{AbacusSummit} emulator grid \citep{maksimova}. While the baseline HOD for the Planck cosmology is optimized for a linear bias of $1.5$ and mean density of $10^{-3}\,h^3\,\mathrm{Mpc}^{-3}$, in alternate cosmologies we modify the target density in a manner motivated by practical observational concerns. In each cosmology, we calculate the true comoving density that would result if an observer assumed the Planck cosmology when constructing a galaxy sample in the redshift range \mbox{$0.2<z<0.45$} with an intended density of $10^{-3}\,h^3\,\mathrm{Mpc}^{-3}$. In a cosmology whose comoving volume in this redshift range is less than that for Planck, for example, a galaxy sample selected in this way will have a true galaxy number density which is greater than $10^{-3}\,h^3\,\mathrm{Mpc}^{-3}$. We also test the effects of changing aperture geometry by halving or doubling the cylinder radii and depths (Sect.~\ref{subsubsec:cics_stoch}) when measuring stochasticity for the baseline HOD in the baseline cosmology.

\begin{figure}
	\includegraphics[width=\columnwidth]{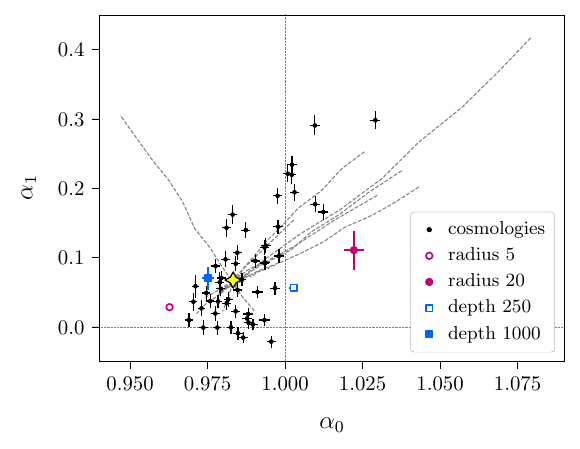}
    \caption{Stochasticity values for baseline HODs derived in alternate cosmologies (black points) and for different cylinder radii (magenta circles) and cylinder depths (blue squares). The tests for alternate cylinder sizes use the baseline HOD in the baseline (Planck 2018) cosmology, and radii and depths in the legend are given in $h^{-1\,}\mathrm{Mpc}$. The 52 non-Planck cosmologies belong to the \textsc{AbacusSummit} emulator grid. All error bars are estimated via spatial jackknife resampling, and errors for radius 5 and depth 250 are of similar scale to the points as plotted. Dashed gray curves correspond to the results from Fig.~\ref{fig:stoch_curves} for reference, and the yellow star at the intersection of the curves indicates the stochasticity of the baseline HOD in the baseline cosmology. Dashed lines at \mbox{$\alpha_0=1$} and \mbox{$\alpha_1=0$} indicate the values corresponding to Poisson shot noise.}
    \label{fig:stoch_cosmo}
\end{figure}

The results of these tests are shown together in Fig.~\ref{fig:stoch_cosmo}. For the baseline HODs derived in different cosmologies, we find that unlike the HOD curves presented so far, a small number of cosmologies produce \mbox{$\alpha_1<0$}, corresponding to an inverse relationship between local matter density and shot noise in galaxy counts. No cosmologies exceed the maximum value of $\alpha_1$ from the HODs in the curves, nor do any produce values of $\alpha_0$ outside the range spanned by the curves. Note again that these results correspond to each cosmology's baseline HOD and that varying the HOD on top of cosmology could amplify or counteract the change in stochasticity due to cosmology alone.

Doubling or halving the depth of the cylinders used to measure galaxy count and matter density has a stronger impact on $\alpha_0$ than on $\alpha_1$, with shallower cylinders producing larger global stochasticity $\alpha_0$ and vice versa. Changing the physical smoothing scale by doubling or halving the cylinder radii affects both stochasticity parameters to a somewhat larger extent. We find that doubling the radius from $10\,h^{-1}\mathrm{Mpc}$ to $20\,h^{-1}\mathrm{Mpc}$ increases both stochasticity parameters and that halving the radius to $5\,h^{-1}\mathrm{Mpc}$ does the opposite. Nevertheless, the values of $\alpha_0$ and $\alpha_1$ produced by these changes in geometry lie within the range spanned by the other tests varying HOD parameters, bias, density, and cosmology. Taking into account the extent of cosmological parameter space spanned by the \textsc{AbacusSummit} emulator cosmologies (5-8$\sigma$ from current CMB plus large-scale structure constraints in any individual parameter, see \citealt{maksimova}), the results in Fig.~\ref{fig:stoch_cosmo} suggest that uncertainty in the details of the galaxy-halo connection, rather than cosmological or geometrical effects, dominate the uncertainty of our prior knowledge on stochasticity in this parametrization.

\subsection{Monte Carlo HOD search}
\label{subsec:mc_hods}
The HODs presented in the preceding sections probe relationships between individual HOD parameters and galaxy stochasticity, and likewise for changes in the target bias or density of a galaxy sample, the underlying cosmology, or the scale and depth of the smoothing filter. However, because these HODs are all linked in some way to the choice of a baseline model, and because they vary a limited number of parameters simultaneously, they necessarily probe a restricted subvolume of the full HOD parameter space. To carry out a more thorough search and to assess the comprehensiveness of the tests presented thus far, we perform a Monte Carlo search for additional HODs, sampling uniformly from a wide volume of parameter space. The parameter ranges for this search are listed in Table~\ref{tab:apdx_mc_param_ranges}. We select the subset of the resulting HODs which are consistent with our target bias and density to within 2 percent while also satisfying the original constraint on the maximum value of $\langle N_\mathrm{g}\rangle$ across all halos (equation~\ref{eq:constraints}). This represents a slightly relaxed constraint relative to the optimization procedure for the previous HODs, all of which achieve the target values of bias and density to within $\sim$0.5 percent. However, given that this degree of uncertainty in bias has only a modest effect on stochasticity (bottom panel of Fig.~\ref{fig:stoch_curves}, where each curve has 11 points total and neighboring points differ in bias by $\sim$1 percent or in density by $\sim$3 percent), we expect that any increase in the range of stochasticity for the Monte Carlo sampled HODs will be dominated by the increase in simultaneous degrees of freedom in the HOD parameters.

\begin{figure*}
    \includegraphics[width=\textwidth]{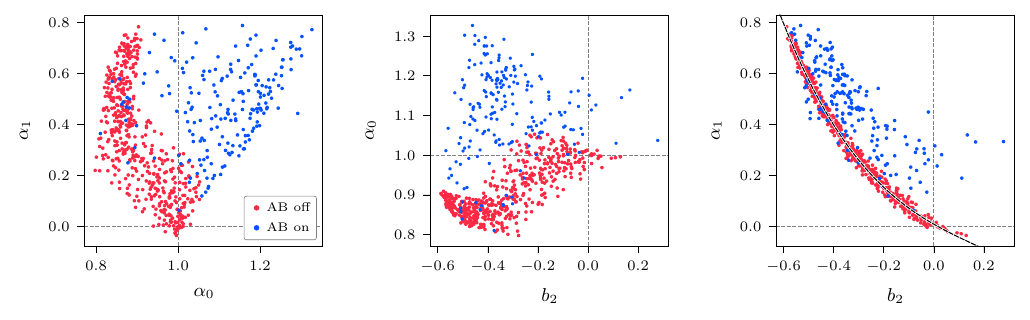}
    \caption{Stochasticity and quadratic bias results for HODs obtained by Monte Carlo sampling from a wide volume of HOD parameter space. Separate searches are performed with assembly bias parameters set to zero (red points, 500 HODs) and with the four assembly bias parameters free (blue points, 200 HODs). Left panel: stochasticity parameters $\alpha_1$ versus $\alpha_0$. Middle panel: $\alpha_0$ versus quadratic bias $b_2$. Right panel: $\alpha_1$ versus $b_2$, showing a tight degeneracy for the HODs without assembly bias (red points). The relation between $b_2$ and $\alpha_1$ for these points is fit by the cubic polynomial \mbox{$\alpha_1(b_2)=-\,1.1\,b_2^{\,3}+0.59\,b_2^{\,2}-0.56\,b_2+0.01$} (dashed black line). Dashed gray lines at \mbox{$\alpha_0=1$}, \mbox{$\alpha_1=0$}, and \mbox{$b_2=0$} indicate the values corresponding to Poisson shot noise and no quadratic bias. The HODs shown are selected based on the constraints of equation~\ref{eq:constraints}, with a 2 percent tolerance on the values of linear bias and density. Note that the red HODs (no assembly bias) are a sample from a lower-dimensional subspace of the parameter volume from which the blue HODs with assembly bias are drawn. Each point represents the mean parameters of 20 galaxy samples generated from a single HOD.}
    \label{fig:stoch_mc}
\end{figure*}

We first sample from the 5-dimensional parameter space for the HOD of equations~\ref{eq:zheng_cen} and \ref{eq:zheng_sat} without assembly bias, allowing the sampler to run until 500 HODs satisfying our constraints have been found. We then expand the search to the 9-dimensional parameter space that includes the 4 assembly bias parameters, for which the search is less efficient and a similar runtime yields 200 HODs meeting the constraints on bias, density, and $\langle N_\mathrm{g}\rangle$. The resulting stochasticity and quadratic bias for the selected HODs are plotted in Fig.~\ref{fig:stoch_mc}, and a corner plot of all selected HOD parameters together with stochasticity is given in Fig.~\ref{fig:apdx_mc_hods}.

In this broader sample of HODs, we find that the overall range of stochasticity is expanded relative to the values spanned by the HOD curves. In particular, the inclusion of assembly bias has a substantial impact on the allowed range of global stochasticity $\alpha_0$. The majority of HODs without assembly bias (red points in Fig.~\ref{fig:stoch_mc}) have \mbox{$\alpha_0<1$}, corresponding to sub-Poisson shot noise in galaxy counts at \mbox{$\delta_\mathrm{m}=0$}, and some achieve slightly negative values of $\alpha_1$ (i.e. an inverse relationship between matter density and shot noise in galaxy counts). With assembly bias turned on, the allowed range of $\alpha_0$ increases at the upper end while the range of $\alpha_1$ remains similar. Another interesting feature of these results is that for an HOD to have a substantial amount of stochasticity overall, that stochasticity must include density dependence ($\alpha_1$). This can be seen in the relative scarcity of HODs with \mbox{$\alpha_0\neq1$} and \mbox{$\alpha_1\simeq0$} in the left panel of Fig.~\ref{fig:stoch_mc}, leading to the sharp V shape in the lower end of the cloud of points. Note that the HODs without assembly bias (red points) are simply a denser sample in a lower-dimensional subspace of the full parameter space that was used to draw the HODs with assembly bias (blue points). 

The majority of the selected HODs produce galaxy catalogs with negative quadratic bias (middle and right panels of Fig.~\ref{fig:stoch_mc}), corresponding to expected galaxy counts in cells that track slightly less than linearly with local matter density. When assembly bias is not present, there is a striking correlation between quadratic bias and $\alpha_1$ (red points in the right panel of Fig.~\ref{fig:stoch_mc}) which is well-described by the cubic polynomial fit \mbox{$\alpha_1(b_2)=-\,1.1\,b_2^{\,3}+0.59\,b_2^{\,2}-0.56\,b_2+0.01$}. The fact that this relation passes very nearly through \mbox{$b_2=\alpha_1=0$} implies that in the absence of assembly bias, nonlinear galaxy bias is required in order to have non-Poisson shot noise, given that $\alpha_0$ is also close to 1 if \mbox{$\alpha_1\simeq0$}. The tightness of this relationship is disrupted when the assembly bias parameters are allowed to vary, as shown by the blue points in the same panel. This potential degeneracy between quadratic bias and the density dependence of stochasticity ($\alpha_1$) has implications for PDF modeling choices, if assembly bias can be constrained for a given galaxy catalog (see Sect.~\ref{subsec:disc_pdf} for further discussion).

Taken together, the full set of 700 Monte Carlo sampled HODs spans stochasticity parameter ranges of \mbox{$\alpha_0\in[0.798,1.327]$} and \mbox{$\alpha_1\in[-0.034,0.788]$}. It is important to emphasize that these results are a test of the range of possible combinations of $\alpha_0$ and $\alpha_1$ in \textsc{redMaGiC}-like galaxy samples and that Fig.~\ref{fig:stoch_mc} should not be interpreted as a probability distribution, as all regions of HOD parameter space are weighted equally during sampling.


\section{Discussion}
\label{sec:disc}
The $\sim$850 HODs and $\sim$30,000 mock galaxy catalogs produced in this work probe a 9-dimensional HOD parameter space (including assembly bias) as well as the additional dimensions of the target bias and number density of the catalogs, the underlying cosmology, and the depth and radius of the smoothing filter. We apply straightforward constraints on overall statistics of a galaxy sample in the form of linear bias, density, and the maximum expected count of galaxies in any halo. We expect that the range of stochasticity produced by our selected HODs would be further reduced with additional constraints on, e.g., the two-point correlation functions of the mock galaxy catalogs. In the combined results of our various tests, stochasticity values for the HODs extend both above and below \mbox{$\alpha_0=1$}, corresponding to super-Poisson and sub-Poisson variance in galaxy counts, respectively, in cells with \mbox{$\delta_\mathrm{m}=0$}. Nearly all of our HODs have \mbox{$\alpha_1>0$}, i.e. a positive relationship between local matter density and stochasticity. We note that this relationship can vary between tracer types; e.g. \citet{Friedrich_pdf} found $\alpha_1$ to be negative for halos above $7.4\times10^{12}\,h^{-1}M_\odot$ in the full-sky light-cone halo catalogs of the Takahashi $N$-body simulation \citep{Takahashi_2017} and for mock galaxies assigned assigned to the same halos by an LRG-like HOD (see figure 9 of \citealt{Friedrich_pdf}).

When varying HOD parameters other than $f_\mathrm{cen}$, we find a generally positive relationship between $\alpha_0$ and $\alpha_1$, as seen in the relevant curves in the first two panels of Fig.~\ref{fig:stoch_curves}. We find the same when varying the target galaxy bias with the HOD fixed at baseline, albeit with greater uncertainty on $\alpha_1$ (bottom panel of Fig.~\ref{fig:stoch_curves}).

Among the stochasticity curves presented in Sect.~\ref{subsec:stoch_results} and Fig.~\ref{fig:stoch_curves}, the fact that the two which vary $f_\mathrm{cen}$ and galaxy density differ from the others in having an inverse relationship between $\alpha_0$ and $\alpha_1$ is noteworthy, as $f_\mathrm{cen}$ and density can be constrained more tightly than many other properties of an HOD or galaxy sample. In the case of a \textsc{redMaGiC} sample, number density is set by construction \citep{Rozo_redmagic}, and $f_\mathrm{cen}$ may be constrained by, e.g., estimates of membership in high-richness clusters (see Sect.~\ref{subsubsec:fiducial_hod}). A tighter upper bound on $f_\mathrm{cen}$ in particular would raise the lower end of the range of $\alpha_0$ among the HOD curves (i.e. the high-$f_\mathrm{cen}$ end of the dashed blue curve in the top panel of Fig.~\ref{fig:stoch_curves}). Well-motivated constraints on the number density and incompleteness ($f_\mathrm{cen}$) of a galaxy catalog may therefore allow for significantly tightened bounds on stochasticity, potentially including the use of non-rectangular prior distributions for $(\alpha_0,\alpha_1)$ to leverage the apparent degeneracy between the two when individual HOD parameters other than $f_\mathrm{cen}$ are varied at fixed density. 

Comparing the effects of changing the HOD (at fixed cosmology) to the effects of changing cosmology (with the non-mass HOD parameters fixed), we find that degrees of freedom in the galaxy-matter connection produce a range of $\alpha_0$ and $\alpha_1$ (left panel of Fig.~\ref{fig:stoch_mc}) that fully covers the range due to changing cosmology, even across the very wide \textsc{AbacusSummit} emulator grid of cosmologies (Fig.~\ref{fig:stoch_cosmo}). This suggests that with galaxy bias and density as constraints, uncertainty in the HOD remains the dominant concern when setting bounds on galaxy stochasticity, provided that the shift in stochasticity when changing cosmology is of similar size at different points in HOD parameter space. When changing cylinder size relative to the baseline radius of $10\,h^{-1}\mathrm{Mpc}$ and depth of $500\,h^{-1}\mathrm{Mpc}$, we find that halving the radius and doubling the depth both decrease $\alpha_0$, despite their opposite effects on the total volume of the cylinder. This may be due to the smaller radius decreasing the fraction of halos which are fully covered by a given cylinder, thus reducing stochasticity as halos hosting multiple galaxies are less likely to be captured in cylinders as a single unit. Increasing the projection depth reduces the overall correlation between pairs of halos in a cylinder, again conceivably reducing stochasticity but via a two-halo effect. As described in Sect.~\ref{subsec:stoch_results_cosmo}, even these twofold changes in the radius or depth of the cylinders produce changes in stochasticity that are subdominant to those arising from the degrees of freedom in the HOD.

\subsection{Assembly bias}
\label{subsec:disc_ab}

A unique feature in the results for the assembly bias parameters $A_\mathrm{cent}$ and $A_\mathrm{sat}$ is the hooked shape of the curves both in HOD parameter space (middle panel of Fig.~\ref{fig:hod_curves}) and in the two stochasticity parameters (middle panel of Fig.~\ref{fig:stoch_curves}). $A_\mathrm{cent}$ and $A_\mathrm{sat}$ control the influence of halo concentration on the effective values of $M_\mathrm{min}$ and $M_1$, respectively, and we find that pushing $A_\mathrm{sat}$ away from zero in either the positive or negative direction produces somewhat similar results in terms of stochasticity, while changing $A_\mathrm{cent}$ within its allowed range has little effect. The observed overall increase in stochasticity regardless of the sign of $A_\mathrm{sat}$ is understandable if the added dependence on concentration simply introduces an additional source of variance in galaxy counts in cells across most or all of the halo mass bins within which concentration is ranked. In contrast, changing either $B_\mathrm{cent}$ or $B_\mathrm{sat}$ produces a more extended, monotonic curve in $M_\mathrm{min}$ versus $M_1$ and in $\alpha_1$ versus $\alpha_0$, with more negative values of either $B$ parameter (i.e. more galaxies in halos with higher environmental density) leading to greater stochasticity (middle panel of Fig.~\ref{fig:stoch_curves}). Conversely, shifting $B_\mathrm{cent}$ to positive values reduces both stochasticity parameters. The decrease in shot noise for \mbox{$B_\mathrm{cent}>0$} implies that introducing a negative correlation between environmental density (ranked within mass bins) and $\langle N_\mathrm{cen}\rangle$ somewhat counters the stochasticity arising from variation in the halo mass function alone, between cells of fixed matter density. It should be noted here that because of the optimization procedure for $M_\mathrm{min}$ and $M_1$, changing the value of an assembly bias parameter will require compensatory changes in the two mass parameters, in order to keep bias and number density fixed. This is expected to be the case e.g. as $B_\mathrm{sat}$ shifts to negative values, preferentially placing satellite galaxies in halos with higher-density environments (i.e. with greater halo bias). It is then reasonable to expect that as $B_\mathrm{sat}$ becomes more negative, $M_1$ must increase to limit satellite galaxy counts in high-bias halos and keep galaxy bias constant, and this is indeed the case for the $B_\mathrm{sat}$ HOD curve in the middle panel of Fig.~\ref{fig:hod_curves}.

Our more thorough Monte Carlo search for \textsc{redMaGiC}-like HODs reveals that the presence or absence of assembly bias has a significant impact on the allowed degree of stochasticity, specifically in the $\alpha_0$ direction (Fig.~\ref{fig:stoch_mc}). Without assembly bias, the basic 5-parameter HOD of equations~\ref{eq:zheng_cen} and \ref{eq:zheng_sat} does produce \mbox{$\alpha_0>1$} to a limited extent, under our constraints on bias and density, but turning on the four additional assembly bias parameters extends the upper bound on $\alpha_0$ from $\sim$$1.05$ to $\sim$$1.33$. In contrast, the allowed range of $\alpha_1$ is effectively spanned by HODs with no assembly bias, indicating that the extent to which shot noise in galaxy counts may depend on local matter density is linked to the manner in which halo occupation depends on halo mass but not concentration or environmental density.

\subsection{The galaxy-matter PDF}
\label{subsec:disc_pdf}

A broader question relevant to this work concerns the parametrization of $P(N_\mathrm{g\,}\vert\,\delta_\mathrm{m})$. The importance of accounting for galaxy stochasticity in both simulations and observations was demonstrated in the DES Year 1 density split statistics analysis \citep{Friedrich_2018,Gruen_2018}, which modeled galaxy bias and stochasticity to linear order in $\delta_\mathrm{m}$. Our results from HODs derived in this work indicate that there exist galaxy-matter connections with \textsc{redMaGiC}-like bias and density for which linear bias alone is not sufficient to accurately describe the PDF of galaxy counts in cells at different fixed matter densities. The comparison between model fits and measured PDFs in Fig.~\ref{fig:pdfs} illustrates the need to model both quadratic bias and density-dependent stochasticity, using the HODs with the two most extreme values of $\alpha_0$ as an example. However, we do find that the range of $\alpha_0$ and $\alpha_1$ across the full set of HODs is similar when fitting stochasticity models with or without quadratic bias (Fig.~\ref{fig:apdx_lin_quad_stoch} and Appendix~\ref{subsec:apdx_lin_quad_stoch}). For the specific purpose of placing conservative priors on stochasticity, then, it may be acceptable to model galaxy bias to linear order only, even if the resulting description of the galaxy-matter PDF is not necessarily accurate for all $\delta_\mathrm{m}$. Regarding the parametrization of stochasticity itself, the results of the Monte Carlo HOD search reinforce the need to model both global stochasticity ($\alpha_0$) and its density dependence ($\alpha_1$) when stochasticity is present in general, as combinations with significant amounts of the former but not the latter appear to be excluded (hence the V-shaped lower boundary in the left panel of Fig.~\ref{fig:stoch_mc}).

A question to be addressed in future work is whether this parametrization can be improved upon, e.g. by identifying degeneracies between bias and stochasticity parameters in order to simplify the model. Our results from the set of Monte Carlo sampled HODs suggest that one such degeneracy exists when assembly bias is negligible, in the form of a tight relationship between quadratic bias $b_2$ and the density dependence of stochasticity $a_1$ (red points in the right panel of Fig.~\ref{fig:stoch_mc}). This implies that for galaxies consistent with our overall constraints, one may model both density-dependent stochasticity and quadratic galaxy bias using only three parameters rather than four, if significant assembly bias can be ruled out. We do not find a similar relationship between any other pair of bias or stochasticity parameters (noting that $b_1$ was tightly constrained by construction). It may, however, be possible to more systematically establish the minimum required dimensionality of the parameter space for $P(N_\mathrm{g\,}\vert\,\delta_\mathrm{m})$ in other ways, such as learning a minimal disentangled representation of the PDF using neural networks.

Even in the absence of tight degeneracies between parameters, other informative features of our results may allow for improved priors on stochasticity. As mentioned in the discussion of the HOD curves in Sec.~\ref{sec:disc}, we find an overall positive correlation between $\alpha_0$ and $\alpha_1$ when varying parameters other than $f_\mathrm{cen}$ and density, opening the possibility of adopting a non-rectangular joint prior on $\alpha_0$ and $\alpha_1$ in a likelihood analysis, if $f_\mathrm{cen}$ and density are well-constrained. Even when all HOD parameters are free, as in the Monte Carlo search of Sec.~\ref{subsec:mc_hods}, we still find excluded regions in the joint distribution of stochasticity parameters, as seen in the left panel of Fig.~\ref{fig:stoch_mc}. In this plot, all \textsc{redMaGiC}-like HODs found in the search lie above a V-shaped boundary that could similarly form the basis for a more informed joint prior on $\alpha_0$ and $\alpha_1$.

\subsection{Impact of constraining galaxy sample properties}

The constraints on bias, density, and the maximum value of $\langle N_\mathrm{g}\rangle$ used to select HODs naturally have an impact on the range of stochasticity parameters measured in the corresponding mock catalogs. The results in this work allow some insight into how our derived bounds on stochasticity would respond if these constraints were further tightened. For the galaxies studied here, the simulation volume and true galaxy density are known. In an observed catalog, in contrast, the true number density is degenerate with the comoving volume of the survey and therefore with the assumed cosmology. When varying the target density over its allowed range of \mbox{$[0.72,1.08]\times10^{-3}\,h^{3}\,\mathrm{Mpc}^{-3}$} (while optimizing $M_1$ and $M_\mathrm{min}$ to keep the other constraints satisfied), we find that the resulting changes in stochasticity (bottom panel of Fig.~\ref{fig:stoch_curves}) are very small relative to the range of stochasticity probed at fixed density when all HOD parameters are free (left panel of Fig.~\ref{fig:stoch_mc}). The effect of changing bias over the range $[1.36,1.54]$ is larger than that for density (bottom panel of Fig.~\ref{fig:stoch_curves}) but still a factor of a few smaller in either $\alpha_0$ or $\alpha_1$ than the range of stochasticity spanned by the degrees of freedom in the HOD itself, when bias is fixed at approximately 1.5. As with cosmology, then, modest uncertainties in bias and number density (which is, as mentioned, degenerate with cosmology) appear to be subdominant to uncertainty in the galaxy-matter connection, for catalogs constrained only by bias, density, and $\max\langle N_\mathrm{g}\rangle$. We therefore do not expect that tightened constraints on bias and density, for example reducing the 2 percent threshold for selecting HODs in the Monte Carlo search, would significantly reduce the allowed ranges of $\alpha_0$ and $\alpha_1$.

In setting a constraint on $\max\langle N_\mathrm{g}\rangle$, the maximum expectation value of the HOD across all halos, we estimated the distribution of \textsc{redMaGiC} galaxy counts in \textsc{redMaPPer} clusters (Fig.~\ref{fig:redmagic}) and set \mbox{$\max\langle N_\mathrm{g}\rangle\in[25,50]$}. Note that because the constraint is only applied to an expectation value, the actual largest count of galaxies in any one halo is still free to be less than 25 or greater than 50 in any individual mock catalog generated by an HOD. Regarding whether tightening this constraint has a meaningful effect on the allowed range of stochasticity, we find that increasing the lower bound on $\max\langle N_\mathrm{g}\rangle$ from 25 to 40 preferentially removes HODs from the central region of the left panel of Fig.~\ref{fig:stoch_mc} and therefore has no significant effect on the overall ranges of $\alpha_0$ and $\alpha_1$. Decreasing the upper bound from 50 to 40 excludes some -- but crucially not all -- HODs in the low-$\alpha_0$, low-$\alpha_1$ region of the same plot and so has little impact on the allowed range of stochasticity as well. Constraining $\max\langle N_\mathrm{g}\rangle$ more tightly than the range used here therefore does little to limit stochasticity when all HOD parameters are free, although it remains possible that imposing e.g. scaling relationships between HOD parameters could make stochasticity more sensitive to the exact bounds on $\max\langle N_\mathrm{g}\rangle$.


\section{Summary and outlook}
\label{sec:concl}

In this work, we use the flexibility of HOD modeling to assess the range of galaxy stochasticity that can plausibly exist for catalogs with \textsc{redMaGiC}-like galaxy bias and density. For galaxies with \mbox{$b\simeq1.5$} and \mbox{$n\simeq10^{-3}\,h^3\,\mathrm{Mpc}^{-3}$}, we obtain a stochasticity range of \mbox{$\alpha_0\in[0.798,1.327]$} and \mbox{$\alpha_1\in[-0.034,0.788]$} in the parametrization of \citet{Friedrich_2018} and \citet{Gruen_2018}. In this parametrization, \mbox{$\alpha_0+\alpha_1\delta_\mathrm{m}\neq1$} corresponds to non-Poisson scatter in galaxy counts at fixed matter overdensity $\delta_\mathrm{m}$. The ranges quoted correspond to a model that includes quadratic galaxy bias, but the two stochasticity parameters remain similar when galaxy bias is modeled to linear order only, as in the original parametrization.

Notably, among the large set of HODs tested, the presence or absence of assembly bias strongly influences the upper bound on global stochasticity ($\alpha_0$). Additionally, we find that for galaxy-matter connections with significant amounts of non-Poisson shot noise, linear galaxy bias alone is insufficient to accurately model mean galaxy count as a function of matter density. The inclusion of quadratic bias in addition to $\alpha_0,\alpha_1$ leads to agreement of the model with the measured PDFs of mock galaxy count (conditioned on matter density) generated by our HODs.

The conservative bounds on stochasticity obtained through this HOD-based approach may be used to motivate priors on $\alpha_0$ and $\alpha_1$ for the purpose of a likelihood analysis which models the galaxy-matter PDF. A companion paper by Ried Guachalla et al. (in prep.) offers a natural application for our results, introducing a methodology for using plausible realizations of nuisance parameters (e.g. the galaxy stochasticity values obtained in this work) to select informative prior distributions for cosmological analyses.

\begin{acknowledgements}

We would like to thank Chun-Hao To, Risa Wechsler, Anik Halder, Yao-Yuan Mao, and members of the Astrophysics, Cosmology, and Artificial Intelligence (ACAI) group at LMU Munich for helpful discussions and feedback.\\

This work was supported by the U.S. Department of Energy through grant DE-SC0013718 and under DE-AC02-76SF00515 to SLAC National Accelerator Laboratory, and by the Kavli Institute for Particle Astrophysics and Cosmology (KIPAC). DB acknowledges support provided by the KIPAC Giddings Fellowship, the Graduate Research \& Internship Program in Germany (operated by The Europe Center at Stanford), the German Academic Exchange Service (Deutscher Akademischer Austauschdienst, DAAD, Short-Term Research Grant 2023 No. 57681230), and the Bavaria California Technology Center (BaCaTec). DG and OF were supported by the Excellence Cluster ORIGINS, which is funded by the Deutsche Forschungsgemeinschaft (DFG, German Research Foundation) under Germany's Excellence Strategy (EXC 2094-390783311). OF was supported by the Fraunhofer-Schwarzschild Fellowship at the Universit\"{a}ts-Sternwarte M\"{u}nchen (LMU Observatory). BRG was funded by the Chilean National Agency for Research and Development (ANID) -- Subdirecci\'{o}n de Capital Humano / Mag\'{i}ster Nacional / 2021 -- ID 22210491 and by the German Academic Exchange Service (DAAD, Short-Term Research Grant 2021 No. 57552337). BRG also gratefully acknowledges support from the Program for Astrophysics Visitor Exchange at Stanford (PAVES).\\

Computing for this project was performed using resources of the National Energy Research Scientific Computing Center (NERSC), a U.S. Department of Energy Office of Science User Facility located at Lawrence Berkeley National Laboratory, operated under Contract No. DE-AC02-05CH11231 using NERSC award HEP-ERCAP0023850.\\

In addition to software cited in the main text, this work made use of \texttt{IPython} \citep{ipython}, \texttt{Matplotlib} \citep{matplotlib}, \texttt{NumPy} \citep{numpy}, \texttt{SciPy} \citep{SciPy}, \texttt{pathos} \citep{pathos}, \texttt{halotools v0.8.1} \citep{Hearin_2017}, and \texttt{hmf} \citep{Murray_hmf}.

\end{acknowledgements}

\section*{Data Availability}

Parameters for the Monte Carlo sampled HODs and an example Python script for populating the \textsc{AbacusSummit} simulations are available at\:\,\mbox{\url{https://github.com/dylan-britt/stochasticity_hods}}.\\The \textsc{AbacusSummit} data products used in this work are made publicly available by the Abacus authors and can be accessed via\:\,\mbox{\url{https://abacusnbody.org}}.


\bibliographystyle{aa}
\bibliography{refs}


\begin{appendix}

\section{}
\label{appendix}

\subsection{Quadratic bias modeling: impact on stochasticity}
\label{subsec:apdx_lin_quad_stoch}

\begin{figure}
    \includegraphics[width=\columnwidth]{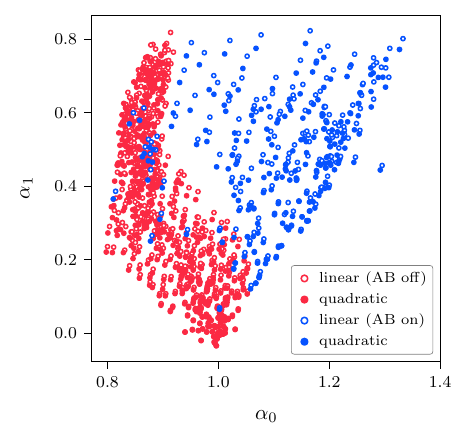}
    \caption{Comparison of stochasticity measurements with and without accounting for quadratic galaxy bias, shown for the full set of Monte Carlo sampled HODs. Plotted are the best fitting values $(\alpha_0,\alpha_1)$ for the stochasticity model of equation~\ref{eq:stoch} with linear bias only (open circles) and for the model of equation~\ref{eq:stoch_quad} that includes quadratic galaxy bias (solid circles). As in Figs.~\ref{fig:stoch_mc} and \ref{fig:apdx_mc_hods}, red points correspond to HODs without assembly bias, and blue points correspond to HODs sampled with the assembly bias parameters free.}
    \label{fig:apdx_lin_quad_stoch}
\end{figure}

To illustrate the impact of quadratic bias modeling on measured stochasticity values, in Fig.~\ref{fig:apdx_lin_quad_stoch} we plot $\alpha_0$ and $\alpha_1$ for the full set of Monte Carlo sampled HODs with and without quadratic bias (equations~\ref{eq:stoch_quad} and \ref{eq:stoch}, respectively). Shifting from the model with quadratic bias to the one with only linear bias causes very little change in the range of $\alpha_0$: \mbox{$[0.798,1.327]\rightarrow[0.800,1.333]$} and a slight increase in the upper end of the range for $\alpha_1$: \mbox{$[-0.034,0.788]\rightarrow[-0.034,0.823]$}.

\subsection{Monte Carlo HOD search}
\label{subsec:appendix_mc}

\begin{table}
    \caption{HOD parameter ranges for the Monte Carlo sampling procedure of Sect.~\ref{subsec:mc_hods}.}
    \label{tab:apdx_mc_param_ranges}
    \begin{center}
        \begin{tabular}{lcc}
            Parameter & Min & Max\\
            \noalign{\smallskip}
            \hline
            \hline
            \noalign{\smallskip}
            $\log_{10}M_\mathrm{min}$ & 11.5 & 12.75\\[0.4em]
            $\log_{10}M_1$ & 12.75 & 15.0\\[0.3em]
            $\sigma_{\log\hspace{-0.1em}M}$ & 0.014 & 1.41\\[0.4em]
            $\alpha_\mathrm{hod}$ & 0.75 & 2.5\\[0.4em]
            $f_\mathrm{cen}$ & 0.05 & 1.0\\[0.4em]
            $A_\mathrm{cent}$ & $-$1.0\phantom{..} & 1.0\\[0.4em]
            $A_\mathrm{sat}$ & $-$1.0\phantom{..} & 1.0\\[0.4em]
            $B_\mathrm{cent}$ & $-$1.0\phantom{..} & 1.0\\[0.4em]
            $B_\mathrm{sat}$ & $-$1.0\phantom{..} & 1.0\\[0.4em]
            \hline
        \end{tabular}
    \end{center}
\end{table}

\begin{figure*}
	\includegraphics[width=\textwidth]{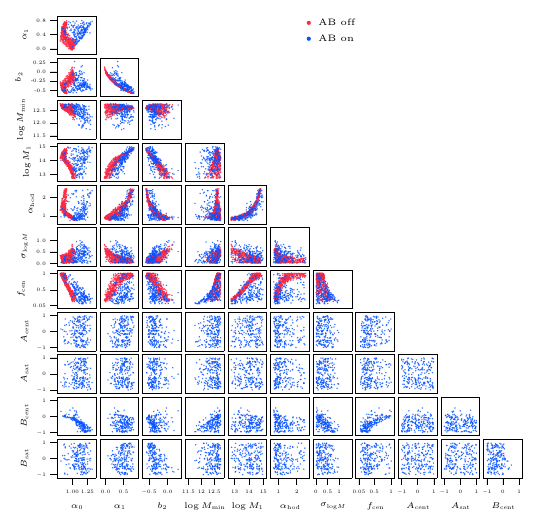}
    \caption{Results of the HOD Monte Carlo sampling procedure described in Sect.~\ref{subsec:mc_hods}. Separate searches were performed with assembly bias (blue points) and without (i.e. all assembly bias parameters set to zero, red points) in the baseline cosmology. The HODs shown achieve the target values of linear galaxy bias ($1.5$) and density ($10^{-3}\,h^3\,\mathrm{Mpc}^{-3}$) to within 2 percent and satisfy the additional constraint that the maximum of $\langle N_\mathrm{g}\rangle$ across all halos falls in the range $[25,50]$ (see equation~\ref{eq:constraints}). The values of stochasticity ($\alpha_0,\alpha_1$) and quadratic bias ($b_2$) shown here are the mean of measurements on 20 mock galaxy catalogs generated by each HOD, as in Fig.~\ref{fig:stoch_mc}. The lower-dimensional search without assembly bias was allowed to run until 500 such HODs were sampled; the search including assembly bias was run until 200 HODs were found. Parameter ranges for the sampling procedure are given in Table~\ref{tab:apdx_mc_param_ranges}.}
    \label{fig:apdx_mc_hods}
\end{figure*}

Table~\ref{tab:apdx_mc_param_ranges} lists the HOD parameter ranges that define the sample space for the Monte Carlo search described in Sect.~\ref{subsec:mc_hods}. Points are drawn uniformly within this volume, and as the search proceeds, various cuts on the parameter space are made to exclude regions which fail to satisfy relaxed versions of our constraints, e.g. failing to achieve the target bias or density to within 50 percent. Fig.~\ref{fig:apdx_mc_hods} shows the selected points in HOD parameter space, together with their quadratic bias and stochasticity. The lower-dimensional search without assembly bias (red points) is more efficient in sampling points which satisfy the constraints and is run until 500 HODs have been found. The higher-dimensional search including the four assembly bias parameters (blue points) is less efficient and is run for a similar total number of samples ($\sim$$10^{\,6}$) until 200 HODs have been found.

\end{appendix}

\end{document}